\documentclass[11pt]{article}
\pdfoutput=1
\usepackage{color}
\usepackage{mhchem}
\usepackage{xcolor}
\usepackage{cite}
\usepackage{hyperref}
\hypersetup{colorlinks=true,linkcolor=red,anchorcolor=black,citecolor=green}
\usepackage[toc,page]{appendix}
\usepackage{mathtools}
\usepackage{amsfonts}
\usepackage{bbold}
\usepackage{textcomp}
\usepackage[DIV13]{typearea}
\usepackage{amsmath, amsthm, amssymb, mathtools,empheq,latexsym,dsfont}
\usepackage{bbm}
\usepackage{amsmath,amssymb,latexsym,dsfont}
\usepackage{slashed, simplewick}
\usepackage[utf8]{inputenc}
\usepackage{graphicx}
\usepackage{graphicx,placeins}
\usepackage{makeidx}
\usepackage[font=small,labelfont=bf]{caption}
\usepackage{nicefrac}
\usepackage{subfigure}
\usepackage{array, bigdelim,multirow,multicol}
\usepackage{fancybox}
\usepackage{bm}
\usepackage{float}
\usepackage{rotating}
\usepackage{colortbl}
\usepackage{booktabs}
\usepackage{nicefrac}
\usepackage[top=2cm,textwidth=16.6cm,textheight=22.75cm]{geometry}
\usepackage{arydshln}
\usepackage{doi}
\usepackage{tikz}
\usetikzlibrary{mindmap,trees}
\usetikzlibrary{decorations.shapes}
\usepackage{framed}
\usepackage[framemethod=TikZ]{mdframed}
\usepackage{enumitem}
\graphicspath{{immagini/}}

\renewcommand{\theequation}{\thesection.\arabic{equation}}

\definecolor{Gray}{gray}{0.92}

\newcommand{\ignore}[1]{}

\newcommand{\be}{\begin{equation}}
\newcommand{\ee}{\end{equation}}
\newcommand{\bea}{\begin{eqnarray}}
\newcommand{\eea}{\end{eqnarray}}

\newcommand{\Id}{\mathbb{1}}

\newcounter{Thm}[section]
\renewcommand{\theThm}{\arabic{section}.\arabic{Thm}}

\newcounter{nodecount}
\newcommand\tabnode[1]{\addtocounter{nodecount}{1} \tikz \node (\arabic{nodecount}) {#1};}

\tikzstyle{every picture}+=[remember picture,baseline]
\tikzstyle{every node}+=[inner sep=0pt,anchor=base,
text depth=.25ex,outer sep=1.5pt]
\tikzstyle{every path}+=[thick, rounded corners]
\tikzset{
	plabel/.style={inner sep=2pt}
}

\makeatletter
\@addtoreset{equation}{section}
\makeatother

\begin{document}

	\title{
		\begin{center}
			
			{\Large\bf  Eclectic flavor group $\Delta(27)\rtimes S_3$ and lepton model building}
	\end{center}}
	\date{}
	\author{
		Cai-Chang Li$^{a,b,c}$\footnote{E-mail: {\tt
				ccli@nwu.edu.cn}},  \
		Gui-Jun Ding$^{d}$\footnote{E-mail: {\tt
				dinggj@ustc.edu.cn}} \
		\\*[20pt]
\centerline{
\begin{minipage}{\linewidth}
\begin{center}
$^a${\it\small School of Physics, Northwest University, Xi'an 710127, China}\\[2mm]
$^b${\it\small Shaanxi Key Laboratory for Theoretical Physics Frontiers, Xi'an 710127, China}\\[2mm]
$^c${\it\small NSFC-SPTP Peng Huanwu Center for Fundamental Theory, Xi'an 710127, China}\\[2mm]
$^d${\it \small Department of Modern Physics, University of Science and Technology of China,\\
Hefei, Anhui 230026, China}\\[2mm]
\end{center}
\end{minipage}}
		\\[10mm]}
	\maketitle
\begin{abstract}
We have performed a systematical study of the eclectic flavor group $\Delta(27)\rtimes S_3$ which is the extension of the traditional flavor symmetry $\Delta(27)$ by the finite modular symmetry $S_3$. Consistency between $\Delta(27)$ and $S_3$ requires that the eight nontrivial singlet representations of $\Delta(27)$ should be arranged into four reducible doublets. The modular transformation matrices are determined for various $\Delta(27)$ multiplets, and the CP-like symmetry compatible with  $\Delta(27)\rtimes S_3$ are discussed. We study the general form of the K\"ahler potential and superpotential invariant under $\Delta(27)\rtimes S_3$, and the corresponding fermion mass matrices are presented. We propose a bottom-up model for lepton masses and mixing based on $\Delta(27)\rtimes S_{3}$, a numerical analysis is performed and the experimental data can be accommodated.

\end{abstract}

\clearpage

\section{Introduction}
	
The standard model (SM) of particle physics based on gauge symmetry gives an excellent description of interactions between the fundamental fermions (quarks and leptons), and it has been precisely tested by a lot of experiments up to TeV scale. The masses of quarks and charged leptons arise from the Yukawa interactions between the charged fermions and Higgs boson in SM, the intergenerational interaction strength is not subject to the constraint of gauge symmetry. Hence the SM can qualitatively explain the fermion masses and flavor mixing but cannot make any precise prediction for their values. The origin of fermion mass hierarchy and flavor mixing is one of most fascinating mysteries of particle physics. The discovery of neutrino oscillation provides new clue to understand this puzzle.
	
Many years of neutrino oscillation experiments have established that neutrinos have tiny masses. The two neutrino mass squared differences and the three lepton mixing angles have been measured with the accuracy of percent level. The mixing pattern of lepton sector is drastically different from that of quark sector. All the three quark mixing angles are small, while the solar and atmospheric neutrino mixing angles are large and the magnitude of the reactor neutrino mixing angle is similar to that of the quark Cabibbo angle~\cite{Workman:2022ynf}. Flavor symmetry acting on the three generations of fermions has been extensively studied to address the pattern of fermion mass hierarchy and mixing angles. In particular, the non-Abelian discrete flavor symmetry could help to naturally explain the large lepton mixing angles~\cite{Altarelli:2005yx,Ishimori:2010au,King:2017guk,Feruglio:2019ybq}.
	
From the top-down perspective, the superstring theory is a promising framework of unifying all four fundamental interactions. The consistency of the theory requires six-dimensional extra compact space besides our lived four-dimensional spacetime. It is remarkable that the non-Abelian discrete flavor symmetry such as $D_4$ and $\Delta(54)$ can arise in certain compactification scheme~\cite{Kobayashi:2004ya,Kobayashi:2006wq,Abe:2009vi}. Moreover, the string duality transformations generate the modular symmetry. The matter fields transform nontrivially under the modular symmetry, consequently the modular symmetry could constrain the flavor structure of quarks and leptons and it enforces the Yukawa couplings to be modular forms~\cite{Feruglio:2017spp}. In bottom-up models with modular symmetry alone, the finite modular groups $\Gamma_N$ and $\Gamma'_N$ ($N=2, 3, 4, 5, 6, 7$) play the role of flavor symmetry, and more generally the finite modular groups can be expressed as the quotient groups of $SL(2, \mathbb{Z})$ over its normal subgroups with finite index~\cite{Liu:2021gwa,Ding:2023ydy}. In the minimal scenario, the complex modulus $\tau$ is the unique source of modular symmetry breaking. One can construct quite predictive models with modular symmetry, and the masses and mixing angles of quarks and leptons can be described in terms of a few free parameters. It is remarkable that the modular symmetry models exhibit a universal behavior in the vicinity of fixed points~\cite{Feruglio:2022koo,Feruglio:2023mii}, independently from details of the models such as the modular weights and representation assignments of matter field under the finite modular groups.  See~\cite{Kobayashi:2023zzc} and references therein for various aspects of modular flavor symmetry. Usually only the minimal K\"ahler potential is adopted in concrete modular models. However, in principle the  K\"ahler potential has many terms compatible with modular symmetry, and they could leads to reduction of the predictability~\cite{Chen:2019ewa,Lu:2019vgm}. How to control the K\"ahler potential is an open question of the modular flavor symmetry approach.

As explained in previous paragraph, the top-down constructions motivated by string theory generally gives rise to both modular symmetry and traditional flavor symmetry. This leads to the idea of eclectic flavor group (EFG) which combines the traditional flavor symmetry with modular symmetry~\cite{Baur:2019kwi,Baur:2019iai,Nilles:2020nnc,Nilles:2020kgo,Nilles:2020tdp,Nilles:2020gvu}. The traditional flavor symmetry and modular symmetry are distinguished  by their action on the modulus $\tau$. The interplay of flavor symmetry and modular symmetry can strongly restrict both the K\"ahler potential and superpotential. The consistency of the theory implies that the mathematical structure of EFG is a semi-direct product $G_{f}\rtimes \Gamma_{N}$ (or $G_{f}\rtimes \Gamma'_{N}$) of the traditional flavor group $G_{f}$ and the finite modular group $\Gamma_{N}$ (or $\Gamma'_{N}$), and each the modular transformation corresponds to an automorphism of traditional flavor symmetry group. In the simplest case that every modular transformation is the trivial identity automorphism of traditional flavor group, the flavor symmetry transformations and modular transformations would be commutable. Then the EFG would reduce to the direct product $G_{f}\times \Gamma_{N}$ (or $G_{f}\times \Gamma'_{N}$), it is the so-called quasi-eclectic flavor group~\cite{Chen:2021prl}, and one can freely choose both traditional flavor symmetry and finite modular group in this case.  The EFG can be consistently combined with CP-like symmetry~\cite{Nilles:2020nnc}, and the corresponding CP transformation has to be compatible with both $G_{f}$ and $\Gamma_{N}$ ($\Gamma'_{N}$). The physical CP transformations are given by the class-inverting automorphisms of the flavor symmetry group~\cite{Chen:2014tpa,Ratz:2019zak}. Therefore one can define a proper generalized CP (gCP) transformation leading to physical CP conservation if a EFG has a class–inverting automorphism. On the other hand, if a EFG do not exhibit class–inverting automorphism, one may still be able to define a CP symmetry when considering a non-generic settings with a special subset of representations which are mapped to their complex conjugates by a CP-like automorphism~\cite{Chen:2014tpa,Ratz:2019zak}. For this kind of EFG, there are usually several possible choices of the CP-like symmetry which leads to distinct physical implications such as the field content and the constraint on the couplings. The EFG has to be broken to obtain realistic fermion masses and mixing angles, consequently both flavon fields and complex modulus $\tau$ are required in EFG models and their vacuum expectation values (VEVs) spontaneously break $G_{f}$ and $\Gamma_{N}$ ($\Gamma'_{N}$) respectively.

EFG is an interesting approach to control the K\"ahler potential. The orbifold $\mathbb{T}^2/\mathbb{Z}_3$ can give rise to the EFG group $\Omega(1)\cong[648,533]$ which is the combination of $\Delta(54)$ flavor symmetry and $T'$ modular symmetry~\cite{Nilles:2020kgo}. Based on the EFG $\Omega(2)\cong[1944, 3448]$ consisting of the traditional flavor group $\Delta(54)$, the finite modular group $T'$ and a $\mathbb{Z}^R_9$ $R$-symmetry,  the first string-derived model was constructed~\cite{Baur:2022hma}. The interplay of flavon alignment and the modulus in the vicinity of modular fixed point leads to naturally protected fermion mass hierarchies, and the flavor observables of both quarks and leptons are reproduced~\cite{Baur:2022hma}. Furthermore, two typical bottom-up models for leptons are constructed with the EFG $\Omega(1)\cong\Delta(27)\rtimes T^\prime\cong[648, 533]$~\cite{Ding:2023ynd}. The experimental data of lepton masses and mixing parameters can be successfully described in terms of five real parameters in the case of gCP symmetry and $\Re \tau=0$ being imposed, and the $\mu-\tau$ reflection symmetry is reproduced exactly.

In the present work, we shall study the EFG $\Delta(27)\rtimes S_{3}\cong[162, 46]$ in a bottom-up way. This EFG is an extension of the traditional flavor group $\Delta(27)$ by the finite modular group $S_{3}$ which is a subgroup of the automorphism group of $\Delta(27)$. The modular transformations $S$ and $T$ which are generators of $S_{3}$ correspond to the outer automorphisms of $\Delta(27)$, while the modular transformations $ST$ and $TS$ correspond to inner automorphisms. In order to consistently combine the modular symmetry $S_{3}$ with traditional flavor symmetry $\Delta(27)$, the eight nontrivial singlet representations of $\Delta(27)$ should be arranged into four doublets $\bm{2_{i}}$ ($i=1,2,3,4$). Considering the modular symmetry further, we find that twelve irreducible two-dimensional representations $\bm{2_{i,m}}$ ($m=0,1,2$) of the EFG $\Delta(27)\rtimes S_{3}$ can be induced from them. Moreover, the three-dimensional irreducible representations $\bm{3}$ and $\bm{\bar{3}}$ of $\Delta(27)$ can be decomposed into a singlet plus a doublet of $S_3$. The matter fields and flavon fields should be assigned to different multiplets of the EFG, and they have definite transformations under $\Delta(27)$ and $S_3$. The superpotential and K\"ahler potential are strongly constrained by $\Delta(27)$ and $S_3$. If the three generations of quark/lepton fields transform as $\bm{3}$ or $\bm{\bar{3}}$ under $\Delta(27)$, the minimal K\"ahler potential is reproduced at leading order. For the singlet plus doublet assignment of matter fields under $\Delta(27)$, the K\"ahler metric is a diagonal yet non-universal matrix at leading order. Consequently normalization of the kinetic terms are expected to give corrections to the flavor observables. Furthermore, we apply the above general results of EFG $\Delta(27)\rtimes S_{3}$ to construct an example  model of lepton masses and mixings.

This paper is organized as follows. We recapitulate the approach of EFG in section~\ref{sec:EFG}, the consistency conditions between $\Delta(27)$ flavor symmetry and $S_3$ modular symmetry are analyzed, and we determine the modular transformation matrices of $S$ and $T$ for different representations of $\Delta(27)$, and the CP-like transformations compatible with $\Delta(27)\rtimes S_{3}$ are fixed. We present the most general form of the K\"ahler potential and superpotential invariant under the EFG $\Delta(27)\rtimes S_{3}$ in section~\ref{sec:mod_general}. We give an example model for neutrino masses and mixing based on the EFG $\Delta(27)\rtimes S_{3}$ in section~\ref{sec:example-model-EFG}. We draw the conclusion and make a summary in section~\ref{sec:conclusion}. The group theory of $\Delta(27)$ and $S_3$ are presented in Appendix~\ref{sec:Delta27_group} and Appendix~\ref{sec:S3_group} respectively. The invariant contractions of two $\Delta(27)$ doublets are given in Appendix~\ref{sec:Delta27-invariant-contractions}. The CP-like transformations compatible with the EFG $\Delta(27)\rtimes S_{3}$ are discussed in Appendix~\ref{sec:EFG_CP_like}.

\section{\label{sec:EFG}Eclectic flavor group $\Delta(27)\rtimes S_{3}$}

The so-called EFG is a nontrivial product of a traditional flavor group $G_{f}$ and a finite modular group $\Gamma_{N}$ ($\Gamma^{\prime}_{N}$), where the finite modular group $\Gamma_{N}$ ($\Gamma^{\prime}_{N}$) is the quotient group of the modular group $\Gamma\cong SL(2, \mathbb{Z})$ over $\pm \Gamma(N)$ ($\Gamma(N)$), and $\Gamma(N)$ is the principal congruence subgroup of level $N$. The full modular group $SL(2, \mathbb{Z})$ is the group of $2\times2$ matrices with integer coefficients and unit determinant,
\begin{equation}
SL(2, \mathbb{Z})=\left\{\begin{pmatrix}
			a  &  b \\
			c  &  d
\end{pmatrix}\Bigg|ad-bc=1, a,b,c,d\in\mathbb{Z}\right\}\,,
\end{equation}
which can be generated by two generators $S$ and $T$ with
\begin{equation}
		S=\begin{pmatrix}
			0  & 1\\-1  & 0
		\end{pmatrix},\qquad T=\begin{pmatrix}
			1 & 1 \\
			0  & 1
		\end{pmatrix}\,.
\end{equation}
The two generators satisfy the multiplication rules
\begin{equation}
		S^4=(ST)^3=\mathbb{1}_2,~~~~S^2T=TS^2\,,
\end{equation}
where $\mathbb{1}_2$ denotes $2\times2$ unit matrix. For a positive integer $N$, the principal congruence subgroup of level $N$ is defined as
\begin{equation}
		\Gamma(N)=\left\{ \gamma \in SL(2,\mathbb{Z}) ~~\Big |~~ \gamma \equiv \begin{pmatrix}
			1 & 0 \\ 0 & 1
		\end{pmatrix} \mod N \right\}\,,
\end{equation}
which implies $T^{N}\in\Gamma(N)$. For $N\leq5$, the multiplication rules of the finite modular group $\Gamma_N\equiv\Gamma/(\pm\Gamma(N))$ and its double covering $\Gamma'_N\equiv\Gamma/\Gamma(N)$ are given by~\cite{Feruglio:2017spp,Liu:2019khw}
\begin{equation}\label{eq:GammaN_defing}
S^{N_{s}}=(ST)^3=T^N=1\,, \qquad S^2T=TS^2\,,
\end{equation}
with $N_s=4$ for $\Gamma'_N$ and $N_s=2$ for $\Gamma_N$, and additional relations are necessary for level $N\geq6$~\cite{deAdelhartToorop:2011re}.
	
Under the action of a traditional flavor transformation $g$ or a modular transformation $\gamma$, the complex modulus $\tau$ and a generic matter field multiplet $\psi$ transform as follow~\cite{Feruglio:2017spp,Liu:2019khw}
\begin{equation}
		\left\{\begin{array}{l}
			\tau \stackrel{g}{\longrightarrow}\tau, \quad \psi\stackrel{g}{\longrightarrow}\rho(g)\psi,~~~~g\in G_f\,, \\
			\tau \stackrel{\gamma}{\longrightarrow}\gamma\tau\equiv\frac{a\tau+b}{c\tau+d},~~~\psi\stackrel{\gamma}{\longrightarrow}(c\tau+d)^{-k_{\psi}}\rho(\gamma)\psi,~~~\gamma=\begin{pmatrix}
				a  &  b\\
				c  & d
			\end{pmatrix}\in\Gamma \,,
		\end{array}\right.
\end{equation}
where $k_{\psi}$ is the modular weight of the matter field multiplet $\psi$, and $\rho(g)$ and $\rho(\gamma)$ are unitary representations of traditional flavor group $G_f$ and the finite modular group $\Gamma_{N}$ or $\Gamma^\prime_{N}$, respectively. Notice that the flavor symmetry transformation leaves the modulus $\tau$ invariant. The modular forms $Y^{(k_Y)}(\tau)$ of level $N$ and weight $k_{Y}$ can be arranged into multiplets of $\Gamma_N$ ($\Gamma^\prime_N$)~\cite{Feruglio:2017spp,Liu:2019khw}:
\begin{equation}
		Y^{(k_Y)}(\tau) \stackrel{\gamma}{\longrightarrow}Y^{(k_Y)}\left(\gamma\tau\right)=\left(c\tau+d\right)^{k_Y}\rho_{Y}(\gamma)\,Y^{(k_Y)}(\tau)\,,
\end{equation}
where $\rho_{Y}(\gamma)$ is a unitary representation of $\Gamma_N$ ($\Gamma^\prime_N$). As the modular multiplet $Y^{(k_Y)}(\tau)$ is holomorphic functions of the complex $\tau$ which is invariant under the action of traditional flavor transformation. Thus, $Y^{(k_Y)}(\tau)$ is invariant under the action of $G_{f}$.
	
In the scheme of EFG, in order to consistently combine a finite modular group with a traditional flavor group, the following consistency condition has to be fulfilled\cite{Nilles:2020nnc,Ding:2023ynd}
\begin{equation}\label{eq:cons_con}
		\rho(\gamma) \rho(g) \rho^{-1}(\gamma)=\rho(u_{\gamma}(g)) \,, \qquad \forall g\in G_{f}\,,
\end{equation}
where $\rho(\gamma)$ represents the automorphism $u_{\gamma}:G_{f}\rightarrow G_{f}$. In the case that $u_{\gamma}$ is the trivial identity automorphism with $u_{\gamma}(g)=g$ for every $\gamma\in \Gamma$ and for all $g\in G_f$, the modular transformation and flavor symmetry transformation would be commutable. This is the so-called quasi-eclectic flavor symmetry~\cite{Chen:2021prl}, and the K\"ahler potential is also constrained by the simultaneous presence of traditional flavor symmetry and modular symmetry. However, the modular symmetry and traditional flavor symmetry can be freely combined together in the quasi-eclectic flavor symmetry, and the resulting models are more complex than these models with either modular symmetry or flavor symmetry alone. Hence we shall be concerned with the case that $u_{\gamma}$ is nontrivial at least for some modular transformation $\gamma$. Then the mathematical structure of the traditional flavor group $G_{f}$ and the finite modular group $\Gamma_{N}$ ($\Gamma^{\prime}_{N}$) is a semi-direct product $G_{f}\rtimes \Gamma_{N}$ ($G_{f}\rtimes \Gamma^{\prime}_{N}$)~\cite{Ding:2023ynd}. It implies that the finite modular group $\Gamma_N$ ($\Gamma^\prime_N$) must be a subgroup of the automorphism group of the traditional flavor group $G_{f}$ and the traditional flavor group $G_{f}$ is a normal subgroup of EFG. In general, the automorphism $u_{\gamma}$ can be outer or inner automorphism of $G_f$ in the scheme of EFG. Here $\rho(g)$ should be the direct sum of all irreducible representations related by the automorphism $u_{\gamma}$, then one could determine the modular transformation $\rho(\gamma)$ by solving the consistency condition of Eq.~\eqref{eq:cons_con}. The resulting $\rho(g)$ and $\rho(\gamma)$ would form a irreducible representation of the EFG, although the restriction to the subgroup $G_{f}$ is usually reducible.
	
As the finite modular groups $\Gamma_N$ and $\Gamma^\prime_{N}$ can be generated by the two generators $S$ and $T$, it is sufficient to impose the consistency condition in Eq.~\eqref{eq:cons_con} on the two outer automorphisms $u_{S}$ and $u_{T}$
\begin{equation}\label{eq:ST_Cons}
		\rho(S)\,\rho(g)\,\rho^{-1}(S)= \rho(u_{S}(g)), \qquad \rho(T)\,\rho(g)\,\rho^{-1}(T)~=~ \rho(u_{T}(g))\,,
\end{equation}
where $\rho(S)$ and $\rho(T)$ are matrix representations of the two automorphisms $u_{S}$ and $u_{T}$ respectively, and they should satisfy the multiplication rules of the finite modular group $\Gamma_N$ or $\Gamma^\prime_N$ in Eq.~\eqref{eq:GammaN_defing}. In other words, the outer automorphisms $u_{S}$ and $u_{T}$ should also satisfy the multiplication rules of the finite modular group $\Gamma_N$ or $\Gamma^\prime_N$:
\begin{equation}\label{eq:uS_uT_rules}
		\left(u_{S}\right)^{N_s} =\left(u_{T}\right)^N =\left(u_{S} \circ u_{T}\right)^3=1,  \qquad \left(u_{S}\right)^2  \circ u_{T} = u_{T} \circ \left(u_{S}\right)^2\;,
\end{equation}
with $N_s=4$ for $\Gamma'_N$ and $N_s=2$ for $\Gamma_N$.
	
\subsection{Traditional flavor group $\Delta(27)$  extended by modular symmetry $\Gamma_{2}\cong S_{3}$}
	
In the present work, we shall consider the EFG extension of the traditional flavor symmetry $\Delta(27)$ by a finite modular group. Following the discussion above, we find that the corresponding finite modular group must be a subgroup of the automorphism group of the traditional flavor group $\Delta(27)$. The group theory of $\Delta(27)$ is discussed in Appendix~\ref{sec:Delta27_group}. The automorphism group of $\Delta(27)$ is $\mathrm{Aut}\left(\Delta(27)\right) \cong [432, 734]$. As the full automorphism group of $\Delta(27)$ only contains two finite modular groups generated by the automorphisms, i.e., $\Gamma_2\cong S_3$  and $\Gamma^\prime_3\cong T^\prime\cong \text{SL}(2,3)$~\cite{Nilles:2020nnc}. Hence the traditional flavor group $\Delta(27)$ can be extended in two ways: by the finite modular groups $\Gamma_2\cong S_{3}$ and $\Gamma_3^\prime\cong T^\prime$ in the case without CP, and the corresponding two EFGs are $\Delta(27)\rtimes S_{3}$  and $\Delta(27)\rtimes T^\prime$, respectively. A comprehensive analysis of EFG models based on $\Omega(1)\cong \Delta(27)\rtimes T^\prime$ is performed in Ref.~\cite{Ding:2023ynd}. In the present work, we are concerned with the traditional flavor group $\Delta(27)$ and its eclectic extension by $\Gamma_2\cong S_{3}$, and the scenarios without/with CP-like symmetry will be studied.
	
The group theory of the finite modular group $\Gamma_2\cong S_{3}$ is given in Appendix~\ref{sec:S3_group}. Two automorphisms that generate the finite modular group $S_3$ can be taken to be~\cite{Nilles:2020nnc}
\begin{equation}\label{eq:S3_uSuT}
		u_S(A)=A^2, \qquad u_T(A)=A^2\,,\qquad u_S(B)=B^2,\qquad u_T(B)=A^2B^2A\,,
\end{equation}
where $A$ and $B$ are the two generators of $\Delta(27)$, please see Eq.~\eqref{eq:D27_Mul_ruls}. Note both automorphisms $u_{S}$ and $u_{T}$ are outer automorphisms of $\Delta(27)$\footnote{One can easy to check that $u_{ST}(A)=A$ and $u_{ST}(B)=ABA^2$. It implies that the automorphism $u_{ST}$ is an inner automorphism of $\Delta(27)$. Analogously $u_{TS}=u^2_{ST}$ is another inner automorphism, and $u_{TST}$ is an outer automorphism with $u_{TST}(A)=A^2$ and $u_{TST}(B)=AB^2A^2$.  }. It is easy to check that the outer automorphisms $u_{S}$ and $u_{T}$ in Eq.~\eqref{eq:S3_uSuT} satisfy the multiplication rules of the finite modular group $S_{3}$
\begin{equation}\label{eq:auto_S3_rules}
		\left(u_{S}\right)^{2} = \left(u_{T}\right)^2=\left(u_{S} \circ u_{T}\right)^3=1\,,
\end{equation}
which can be obtained from  Eq.~\eqref{eq:uS_uT_rules} by taking $N_s=2$ and $N=2$. If one consider the outer automorphisms $u_{S}$ and $u_{T}$ in Eq.~\eqref{eq:S3_uSuT}, the traditional flavor group $\Delta(27)$ shall be extended to the EFG $\Delta(27)\rtimes S_{3}$.
	
In order to determine the explicit expressions of the modular transformations $\rho(S)$ and $\rho(T)$  corresponding to the outer automorphisms $u_{S}$ and $u_{T}$, we should analyse how the two outer automorphisms $u_{S}$ and $u_{T}$ act on the conjugacy classes and irreducible representations of $\Delta(27)$.  From the conjugacy classes of $\Delta(27)$ in Eq.~\eqref{eq:D27CC}, we see that the outer automorphisms $u_S:(A,\,B)\rightarrow (A^2,\,B^2)$ and $u_T:(A,\,B)\rightarrow (A^2,\,A^2B^2A)$ act on all conjugacy classes as follows
\begin{eqnarray}
		\nonumber u_S,~ u_T~: && 1C_1\leftrightarrow 1C_1, \qquad3C^{(1)}_3\leftrightarrow 3C^{(5)}_3, \qquad  3C^{(2)}_3\leftrightarrow 3C^{(4)}_3,
		\qquad 3C^{(3)}_3\leftrightarrow 3C^{(6)}_3, \\
		\label{eq:uS_CC} &&  3C^{(7)}_3\leftrightarrow 3C^{(8)}_3, \qquad 1C^{(1)}_3\leftrightarrow 1C^{(1)}_3, \qquad   1C^{(2)}_3\leftrightarrow 1C^{(2)}_3\,,
\end{eqnarray}
which is displayed in table~\ref{tab:character_Delta27}.

As we know, the outer automorphisms of a group not only map one conjugacy class to another but also map one irreducible representation to another, while the character table is invariant. The consistency condition Eq.~\eqref{eq:cons_con} may be understood as the action of an outer automorphism $u_{\gamma}$ on a representation $\rho$ of traditional flavor symmetry as follow
\begin{equation}\label{eq:rep_trans}
		u_{\gamma}:\rho\rightarrow\rho^\prime=\rho\circ u_{\gamma}\,,
\end{equation}
where $\rho$ could be any reducible or irreducible representation of $\Delta(27)$.  The consistency condition \eqref{eq:cons_con} requires that $\rho$ and $\rho'$ should be equivalent representations and the modular transformation $\rho(\gamma)$ is the similarity transformation. The solution for $\rho(\gamma)$ exists if and only if  $\rho$ contains all those irreducible representations of $\Delta(27)$ related by the outer automorphism $u_{\gamma}$ as Eq.~\eqref{eq:rep_trans}. In fact, $\rho$ would be the restriction of certain representation of the EFG $\Delta(27)\rtimes S_{3}$ on the flavor symmetry group $\Delta(27)$.

Then we proceed to consider the actions of outer automorphisms $u_{S}$ and $u_T$ on all irreducible representations of $\Delta(27)$.  It is straightforward to verify that both of the two outer automorphisms $u_S$ and $u_T$ act on the eight nontrivial singlet irreducible representations $\bm{1_{r,s}}$ of $\Delta(27)$ with $\bm{r,s}\neq0$ as
\begin{equation}\label{eq:uS_uT_singlets}
		u_{S},~u_{T}~:~ \bm{1_{0,1}}\leftrightarrow \bm{1_{0,2}}\,,\quad  \bm{1_{1,0}}\leftrightarrow \bm{1_{2,0}}\,, \quad \bm{1_{1,1}}\leftrightarrow \bm{1_{2,2}}\,, \quad  \bm{1_{1,2}}\leftrightarrow \bm{1_{2,1}}\,,
\end{equation}
which indicates that each one of the eight nontrivial one-dimensional representations is related to another one by the $S_{3}$ modular symmetry. As a consequence, consistency between the modular symmetry $S_3$ and flavor symmetry $\Delta(27)$ requires that the eight non-trivial singlet representations of $\Delta(27)$ should be arranged into the following four reducible doublets of $\Delta(27)$:
\begin{equation}
\label{eq:4-red-doublet-D27}\bm{2_1}\equiv(\bm{1_{0,1}}, \bm{1_{0,2}})^T, \quad \bm{2_2}\equiv(\bm{1_{1,1}}, \bm{1_{2,2}})^T, \quad
\bm{2_3}\equiv(\bm{1_{1,0}}, \bm{1_{2,0}})^T, \quad
\bm{2_4}\equiv(\bm{1_{1,2}}, \bm{1_{2,1}})^T\,.
\end{equation}
The representation matrices of the $\Delta(27)$ generators $A$ and $B$ read off as
\begin{eqnarray}\label{eq:Delta27_2D_Reps}
\nonumber &&\bm{2_{1}}:~~\rho_{\bm{2_{1}}}(A)=\text{diag}(1,1), \qquad \rho_{\bm{2_{1}}}(B)=\text{diag}(\omega,\omega^2 )\,, \\
\nonumber &&\bm{2_{2}}:~~\rho_{\bm{2_{2}}}(A)=\text{diag}(\omega,\omega^2 ),\qquad \rho_{\bm{2_{2}}}(B)=\text{diag}(\omega,\omega^2 )\,, \\
\nonumber &&\bm{2_{3}}:~~\rho_{\bm{2_{3}}}(A)=\text{diag}(\omega,\omega^2 ),  \qquad \rho_{\bm{2_{3}}}(B)=\text{diag}(1,1)\,, \\
&&\bm{2_{4}}:~~\rho_{\bm{2_{4}}}(A)=\text{diag}(\omega,\omega^2 ),\qquad \rho_{\bm{2_{4}}}(B)=\text{diag}(\omega^2, \omega )\,.
\end{eqnarray}
Analogously we find that the trivial singlet representation $\bm{1_{0,0}}$, and the two three-dimensional representations $\bm{3}$ and $\bm{\bar{3}}$ of $\Delta(27)$ are all invariant under the actions of $u_{S}$ and $u_{T}$, i.e.
\begin{equation}
u_{S},~u_{T}~:~ \bm{1_{0,0}}\rightarrow \bm{1_{0,0}}\,,\quad  \bm{3}\rightarrow \bm{3},\quad \bm{\bar{3}}\rightarrow \bm{\bar{3}}\,.
\end{equation}
Hence the three irreducible representations $\bm{1_{0,0}}$, $\bm{3}$ and $\bm{\bar{3}}$ need not be extended to include other irreducible representations of $\Delta(27)$. We summarize the actions of the automorphisms $u_{S}$ and $u_{T}$ on the irreducible representations of $\Delta(27)$ in table~\ref{tab:character_Delta27}. Accordingly the modular transformations of $S$ and $T$ are fixed by the consistency condition in Eq.~\eqref{eq:ST_Cons} for $g=A, B$, i.e.,
\begin{eqnarray}
\nonumber &&\rho_{\bm{r}}({S})\,\rho_{\bm{r}}(A)\,\rho^{-1}_{\bm{r}}({S}) ~=~ \rho_{\bm{r}}(A^2),\qquad \rho_{\bm{r}}({T})\,\rho_{\bm{r}}(A)\,\rho^{-1}_{\bm{r}}({T}) ~=~ \rho_{\bm{r}}(A^2)\,,\\
\label{eq:cons_D27_S3}&&\rho_{\bm{r}}({S})\,\rho_{\bm{r}}(B)\,\rho^{-1}_{\bm{r}}({S}) ~=~ \rho_{\bm{r}}(B^2), \qquad \rho_{\bm{r}}({T})\,\rho_{\bm{r}}(B)\,\rho^{-1}_{\bm{r}}({T}) ~=~ \rho_{\bm{r}}(A^2B^2A)\,,
\end{eqnarray}
where $\bm{r}$ can be the three irreducible representations  $\bm{1_{0,0}}$, $\bm{3}$ and $\bm{\bar{3}}$,  and the four reducible two-dimensional representations $\bm{2_{i}}$ ($i=1,2,3,4$) of $\Delta(27)$. Furthermore, as elements $S$ and $T$ are the generators of the finite modular group $S_{3}$, the modular transformations $\rho_{\bm{r}}(S)$ and $\rho_{\bm{r}}({T})$ have to satisfy the multiplication rules of the finite modular group $S_{3}$:
\begin{equation}\label{eq:S3_mul_rules_rhoST}
		\rho^2_{\bm{r}}(S)=\rho^2_{\bm{r}}(T)=\rho^3_{\bm{r}}(ST)=\mathbb{1}_{\bm{r}}\,.
\end{equation}
For the trivial singlet $\bm{r}=\bm{1_{0,0}}$, it is easy to check that the solutions for $\rho_{\bm{r}}(S)$ and $\rho_{\bm{r}}(T)$ are the two one-dimensional representations of the finite modular group $S_{3}$, i.e.
\begin{equation}
		\rho_{\bm{1^{a}_{0, 0}}}(S)=(-1)^a, \qquad \rho_{\bm{1^{a}_{0, 0}}}(T)=(-1)^a \,,
\end{equation}
where $a=0,1$.
	
For the triplet representation $\bm{3}$ of $\Delta(27)$, the consistency condition and multiplication rules of the finite modular group $S_{3}$ fix the modular transformations to be:
\begin{equation}\label{eq:3DS3fromD27matrices}
		\rho_{{\bm {3^{a}}}}({S}) =(-1)^{a}\left(
		\begin{array}{ccc}
			1 & 0 & 0 \\
			0 & 0 & 1 \\
			0 & 1 & 0 \\
		\end{array}
		\right) \;,\quad
		\rho_{{\bm{3^a}}}({T}) =(-1)^{a}\left(
		\begin{array}{ccc}
			0 & 1 & 0 \\
			1 & 0 & 0 \\
			0 & 0 & 1 \\
		\end{array}
		\right) \;,
\end{equation}
with $a=0,1$.  From the three-dimensional matrices $\rho_{\bm{3}}({A})$, $\rho_{\bm{3}}({B})$ in Eq.~\eqref{eq:Delta27_irre}, one can directly obtain
\begin{equation}\label{eq:3Rep_AST}
		\rho_{{\bm {3^{a}}}}(ST)=\rho_{{\bm{ 3}}}({A})\,.
\end{equation}
The matrices $\rho_{{\bm{ 3}}}({A})$, $\rho_{{\bm{ 3}}}({B})$, $\rho_{{\bm{ 3^{a}}}}({S})$ and $\rho_{{\bm{ 3^{a}}}}({T})$ generate a EFG $\Delta(54)\cong [54,8]$. The three-dimensional representation in Eq.~\eqref{eq:3DS3fromD27matrices} is a reducible representation of $S_{3}$ and it is the direct sum of a singlet and a doublet representations of $S_{3}$,
\begin{equation}
		\bm{3^a}=\bm{1^a} \oplus\bm{2}\,.
\end{equation}
For a triplet $\Phi_{\bm{3^{a}}}=(\phi_{1},\phi_{2},\phi_{3})^T$ transforming as $\rho_{\bm{3^a}}$ will decompose to one singlet $\bm{1^a}$ and one doublet $\bm{2}$ of $S_{3}$ as follow
\begin{equation}\label{eq:triplet-to-siglet-doublete}
		\bm{1^{a}}~:~\frac{1}{\sqrt{3}}(\phi_{1}+\phi_{2}+\phi_{3}), \qquad  \qquad \bm{2}~:~\frac{P^{a}_{2}}{\sqrt{6}}\left(\begin{array}{c}\phi_{1}+\phi_{2}-2  \phi_{3}\\
			\sqrt{3}(\phi_{2}-\phi_{1})\end{array}\right)\,,
\end{equation}
with
\begin{equation}
		P_{2}=\left(\begin{array}{cc}0&-1\\
			1 & 0\end{array}\right)\,.
\end{equation}
For the representation $\bm{\bar{3}}$ of $\Delta(27)$, the corresponding modular transformations $\rho_{\bm{\bar{3}^{a}}}({S})$ and $\rho_{\bm{\bar{3}^{a}}}({T})$ coincide with $\rho_{\bm{3^{a}}}({S})$ and $\rho_{\bm{3^{a}}}({T})$ in Eq.~\eqref{eq:3DS3fromD27matrices} respectively. As a consequence, an $S_{3}$ triplet in $\bm{\bar{3}^a}$ can be decomposed into a singlet and a doublet of $S_3$ modular symmetry, as shown in Eq.~\eqref{eq:triplet-to-siglet-doublete}.

For each one reducible two-dimensional representation $\bm{2_{i}}$, there are three independent solutions for the modular transformations  that fulfils  both the consistency condition~\eqref{eq:cons_D27_S3} and the multiplication rules~\eqref{eq:S3_mul_rules_rhoST}. The representation matrices of the two generators $S$ and $T$ are denoted by $\rho_{\bm{2_{i,m}}}({S})$ and $\rho_{\bm{2_{i,m}}}({T})$ respectively with $m=0,1,2$. In our working basis, they are determined to be
\begin{equation}\label{eq:2DS3fromD27matrices}
		\rho_{\bm{ 2_{i,m}}}({S}) = \left(
		\begin{array}{cc}
			0 ~& 1 \\
			1 ~& 0
		\end{array}\right)\;, \qquad
		\rho_{\bm{ 2_{i,m}}}({T}) = \left(\begin{array}{cc}	0 & \omega^m \\
			\omega^{-m} & 0 \end{array}\right) \,.
\end{equation}
We find that the matrices  $\rho_{{\bm {2_{i}}}}({A})$, $\rho_{{\bm {2_{i}}}}({B})$, $\rho_{{\bm {2_{i,m}}}}({S})$ and $\rho_{{\bm {2_{i,m}}}}({T})$ expand into a matrix group of $S_3$. Similar to Eq.~\eqref{eq:3Rep_AST} for the triplet representation, there is at least one solution of $\rho_{{\bm {2_{i,m}}}}({ST})$ identical with the representation matrix $\rho_{{\bm {2_{i}}}}({A})$ for each two-dimensional representation $\bm{2_{i}}$, i.e.
\begin{equation}
		\label{eq:ST-eq-A}\rho_{\bm{ 2_{i,m}}}(ST)=\rho_{\bm{2_i}}(A), ~~\text{for}~~ (i,m)=(1,0),(2,2),(3,2),(4,2)\,,
\end{equation}
while\footnote{For the direct sum of $\bm{ 3^{a}}\oplus\bm{ 2_{i,m}}$ with the values of the indices $i, m$ in Eq.~\eqref{eq:ST-neq-A}, the representation matrices of $A$, $B$, $S$, $T$ generate a matrix group isomorphic to the EFG $\Delta(27)\rtimes S_3$.}
\begin{equation}
		\label{eq:ST-neq-A}\rho_{\bm{ 2_{i,m}}}(ST)\neq\rho_{\bm{2_i}}(A), ~~\text{for}~~ (i,m)=(1,1), (1,2), (2,0), (2,1), (3,0), (3,1), (4,0), (4,1)\,.
\end{equation}
Since $u_{ST}$ is an inner automorphism of $\Delta(27)$, consequently the modular transformation $\rho({ST})$ must coincide with certain flavor symmetry transformation $\rho(g)$ with $g=A\in\Delta(27)$, as shown in Eq.~(\ref{eq:3Rep_AST},\ref{eq:ST-eq-A}). Then the modular forms in the Yukawa couplings must be invariant under the action of $ST$, consequently only modular form singlets of $S_3$ are allowed and they can absorbed into the coupling constants. For the solutions of $\rho_{\bm{ 2_{i,m}}}(ST)\neq\rho_{\bm{2_i}}(A)$ in Eq.~\eqref{eq:ST-neq-A}, modular form doublets of $S_3$ can enter into the Yukawa couplings and this provides intriguing possibility for model building. Moreover, the representations $\bm{2_{i,m}}$ can be decomposed into $\bm{1} \oplus\bm{1^\prime}$ of $S_3$  for $m=0$ while it is equivalent to the doublet representation $\bm{2}$ for $m=1, 2$. For the doublet fields $\Phi_{\bm{2_{i,m}}}=(\phi_{1},\phi_{2})^T$, we find the following decomposition
\begin{eqnarray}
		\nonumber && \Phi_{\bm{2_{i,0}}}~:~\frac{1}{\sqrt{2}}\left(\phi_{1}+\phi_{2}\right)\sim \bm{1},\qquad \qquad \frac{1}{\sqrt{2}}\left(\phi_{1}-\phi_{2}\right)\sim\bm{1^{\prime}}\,, \\
		\nonumber && \Phi_{\bm{2_{i,1}}} ~:~ \frac{1}{\sqrt{2}}\left(\begin{array}{c} \phi_{1}+\omega \phi_{2}\\-i ( \phi_{1}-\omega\phi_{2})\end{array}\right)~\sim~ \bm{2} \,,\\
		&& \Phi_{\bm{2_{i,2}}} ~:~  \frac{1}{\sqrt{2}}\left(\begin{array}{c} \phi_{1}+\omega^2 \phi_{2}\\i (\phi_{1}-\omega^2\phi_{2})\end{array}\right)~\sim~ \bm{2}\,.
\end{eqnarray}
In short, from the representations of $\Delta(27)$ flavor symmetry we have reached the irreducible multiplets of the EFG $\Delta(27)\rtimes S_{3}$ denoted as $\bm{1^{a}_{0, 0}}$, $\bm{3^a}$, $\bm{\bar{3}^a}$, $\bm {2_{i,m}}$ with $a=0,1$, $i=1, 2, 3, 4$, $m=0,1,2$. Another irreducible two-dimensional representation of the EFG can be induced from the $S_{3}$ irreducible representation $\bm{2}$ and in the following it is labelled as $\bm{2_{0}}$, in which the representation matrices of generators $A$, $B$, $S$ and $T$ are
\begin{equation}
		\rho_{\bm{2_{0}}}(A)=\mathbb{1}_{2}, \quad \rho_{\bm{2_{0}}}(B)=\mathbb{1}_{2}, \quad \rho_{\bm{2_{0}}}(S)=\rho_{\bm{2}}(S), \quad \rho_{\bm{2_{0}}}(T)=\rho_{\bm{2}}(T)\,,
\end{equation}
where the representation matrices $\rho_{\bm{2}}(S)$ and $\rho_{\bm{2}}(T)$ are given by Eq.~\eqref{eq:Tp_irre}. Furthermore the remaining two irreducible representations of EFG $\Delta(27)\rtimes S_{3}$ are of dimension six and they are given by~\cite{semidireci_Reps}
\begin{equation}
		\bm{6}=\bm{3^0}\otimes\bm{2_{0}}\cong\bm{3^1}\otimes\bm{2_{0}}, \qquad
		\bm{\bar{6}}=\bm{\bar{3}^0}\otimes\bm{2_{0}}\cong\bm{\bar{3}^1}\otimes\bm{2_{0}}\,,
\end{equation}
where the operation $\otimes$ denotes as Kronecker product of matrix. Then one can easily write out the six-dimensional representation matrices of the EFG generators $A$, $B$, $S$ and $T$ from the representations $\bm{3^0}$, $\bm{\bar{3}^0}$ and $\bm{2_{0}}$. Thus, all twenty-one irreducible representations of the EFG $\Delta(27)\rtimes S_{3}\cong[162,46]$ are already found, and the representations matrices of the generators $A$, $B$, $S$ and $T$ are summarized in table~\ref{tab:EFG_D27S3_Reps}.
Furthermore, we see that the representations $\bm{1^{a}_{0,0}}$ and $\bm{2_{0}}$ are real, the doublet representations $\bm{2_{i,m}}$ are not real but self-conjugate, and the triplet representations $\bm{3^{a}}$ and the sextet representation $\bm{6}$ are the complex conjugate of $\bm{\bar{3}^{a}}$ and $\bm{\bar{6}}$ respectively.

\begin{table}[t!]
\begin{center}
\renewcommand{\tabcolsep}{2.5mm}
\renewcommand{\arraystretch}{1.1}
\begin{tabular}{|c|c|c|c|c|c|}\hline\hline
\multicolumn{5}{|c|}{The irreducible representations of the EFG $\Delta(27)\rtimes S_{3} \cong [162,46]$} \\ \hline \hline

~~  &  $A$  &   $B$  &  $S$  &   $T$ \\ \hline
		
$\bm{1^{a}_{0,0}}$ & $1$   &  $1$ &$(-1)^a$   &  $(-1)^a$    \\ \hline

$\bm{2_{0}}$ & $\mathbb{1}_{2}$ & $\mathbb{1}_{2} $
			&  $\rho_{\bm{2}}(S)$
			&  $\rho_{\bm{2}}(T)$	\\ \hline

$\bm{2_{i,m}}$ &  $\rho_{\bm{2_{i}}}(A)$ 	& $\rho_{\bm{2_{i}}}(B)$
			& $\rho_{\bm{2_{i,m}}}(S) $  & $\rho_{\bm{2_{i,m}}}(T)$   \\ [0.02in]\hline

$\bm{3^{a}}$ &  $\rho_{\bm{3}}(A)$ 	& $\rho_{\bm{3}}(B)$
			& $\rho_{\bm{3^{a}}}(S) $  & $\rho_{\bm{3^{a}}}(T)$   \\ [0.02in]\hline			
			
$\bm{\bar{3}^{a}}$ &  $\rho_{\bm{\bar{3}}}(A)$ 	& $\rho_{\bm{\bar{3}}}(B)$
			& $\rho_{\bm{\bar{3}^{a}}}(S) $ & $\rho_{\bm{\bar{3}^{a}}}(T)$   \\ [0.02in]\hline

$\bm{6}=\bm{3^{0}}\otimes\bm{2_{0}}$ & $\rho_{\bm{3}}(A)\otimes \mathbb{1}_{2}$ & $\rho_{\bm{3}}(B)\otimes \mathbb{1}_{2}$
			& $\rho_{\bm{3^{0}}}(S)\otimes \rho_{\bm{2}}(S)$
			& $\rho_{\bm{3^{0}}}(T)\otimes \rho_{\bm{2}}(T)$\\[0.050in] \hline		
			
$\bm{\bar{6}}=\bm{\bar{3}^{0}}\otimes\bm{2_{0}}$ & $\rho_{\bm{\bar{3}}}(A)\otimes \mathbb{1}_{2}$
			& $\rho_{\bm{\bar{3}}}(B)\otimes \mathbb{1}_{2}$
			& $\rho_{\bm{\bar{3}^{0}}}(S)\otimes \rho_{\bm{2}}(S)$
			& $\rho_{\bm{\bar{3}^{0}}}(T)\otimes \rho_{\bm{2}}(T)$\\[0.050in] \hline		
			\hline
\end{tabular}
\caption{\label{tab:EFG_D27S3_Reps}The irreducible representations of the EFG $\Delta(27)\rtimes S_{3} \cong [162,46]$ and the representation matrices of the generators $A$, $B$, $S$ and $T$ in our working basis, where $\omega=e^{2\pi i/3}$, the indices $a=0,1$, $m=0,1,2$ and $i=1, 2, 3, 4$.  }
\end{center}
\end{table}

\subsection{\label{subsec:gCP}Including CP-like symmetry}

\begin{figure}[t!]
\centering
\includegraphics[width=0.85\linewidth]{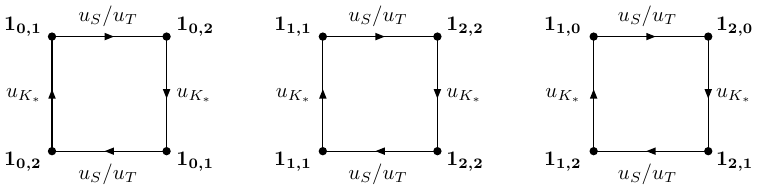}
\caption{\label{fig:action-uS-uT-uKstar} The action of the automorphisms $u_S$, $u_T$ and $u_{K_*}$ on the non-trivial singlet representations of $\Delta(27)$.}
\end{figure}

One can combine the EFG $G_{f}\rtimes \Gamma_{N}$ ($G_{f}\rtimes \Gamma^{\prime}_{N}$) with the CP-like symmetry by introducing a new generator $K_{*}$~\cite{Nilles:2020nnc}, which corresponds to an automorphism of both the traditional flavor symmetry $G_{f}$ and the finite modular group $\Gamma_{N}$ ($\Gamma^{\prime}_{N}$). The CP-like transformation $K_{*}$ acts on the modulus $\tau$, the matter field and the modular form multiplets of level $N$ and weight $k_{Y}$ as follows,
\begin{equation}\label{eq:K*_to_matter}
		\tau\stackrel{K_{*}}{\longrightarrow}-\bar{\tau},\quad	\psi(x)\stackrel{K_{*}}{\longrightarrow} \rho(K_*)[\psi^{\dagger}(t, -{\bm x})]^T, \quad Y^{(k_{Y})}(\tau)\stackrel{K_{*}}{\longrightarrow} Y^{(k_{Y})}(-\bar{\tau})= \rho(K_*)(Y^{(k_{Y})}(\tau))^{*}\,,
\end{equation}
where the CP-like transformation $\rho(K_*)$ is a unitary matrix, and the obvious action of CP on the spinor indices is omitted for the case of $\psi$ being spinor. Requiring that the gCP transformation $K_{*}$ be of order 2 with $(K_{*})^2=1$, we can obtain
\begin{equation}\label{eq:rho_K*_order2}
		\rho^*(K_*)= \rho^{-1}(K_*)\,.
\end{equation}
The CP-like transformation has to be compatible with both the traditional flavor symmetry and the finite modular group, and its allowed form is strongly constrained by the corresponding restricted consistency conditions. The consistency between the EFG and  CP-like symmetry requires the following consistency conditions have to be satisfied~\cite{Novichkov:2019sqv,Ding:2021iqp}:
\begin{eqnarray}
\label{eq:con_uK} &\rho(K_*)\rho^*(g)\rho^{-1} (K_*)= \rho(u_{{K}_*}(g))\,, \qquad \forall g\in G_{f}\,,\\
\label{eq:consistency_K*_mod} &\rho(K_*)\rho^{*}(S)\rho^{-1}(K_*)=\rho^{-1}(S),\qquad 	\rho(K_*)\rho^{*}(T)\rho^{-1}(K_*)=\rho^{-1}(T)\,,
\end{eqnarray}
where $u_{{K}_*}$ is an automorphism of the traditional flavor symmetry group $G_f$. It is sufficient to consider the element $g$ being the generators of $G_f$, and one can fix the explicit form of the CP-like transformation $\rho(K_*)$ up to an overall irrelevant phase by solving Eqs.~(\ref{eq:con_uK},\ref{eq:consistency_K*_mod}). Hence the automorphism $u_{K_*}$ of the traditional flavor group $G_f$ should satisfy the following relations:
\begin{equation}\label{eq:auto_Kstar_rules}
(u_{{K}_*})^2=1, \qquad  u_{{K}_*}\circ u_{S}\circ u_{{K}_*}=u_{{S}}^{-1}, \qquad  u_{{K}_*}\circ u_{T}\circ u_{{K}_*}=u_{{T}}^{-1}\,.
\end{equation}
As regards the concerned EFG $\Delta(27)\rtimes S_3$, the actions of the automorphism $u_{K_*}$ on the $\Delta(27)$ generators $A$ and $B$ can be taken to be~\cite{Nilles:2020nnc}
\begin{equation}\label{eq:CPfromDelta27}
u_{{K}_*}(A) ~=~ A\,,\qquad u_{{K}_*}(B) ~=~ A\, B^2\,A\,,
\end{equation}
which implies that $u_{K_*}$ is an outer automorphism of $\Delta(27)$. Notice that the choice for the automorphism of the CP-like symmetry is not unique in the EFG $\Delta(27)\rtimes S_3$, see Appendix~\ref{sec:EFG_CP_like} for details. If the above CP-like symmetry is imposed, the EFG is extended and the ID of the enlarged group is [324, 121] in GAP~\cite{GAP,SmallGroups}, and we need to discuss the actions of the outer automorphism $u_{K_{*}}$ on the conjugacy classes and the irreducible representations of $\Delta(27)$. The outer automorphism $u_{K_*}$ acts on the conjugacy classes as
\begin{equation}
u_{K_*}:~1C^{(1)}_3\leftrightarrow 1C^{(2)}_3, \quad 3C^{(1)}_3\leftrightarrow 3C^{(8)}_3, \quad  3C^{(2)}_3\leftrightarrow 3C^{(4)}_3, \quad 3C^{(5)}_3\leftrightarrow 3C^{(7)}_3\,.
\end{equation}
The remaining three conjugacy classes of $\Delta(27)$ are invariant under $u_{K_*}$. Then we shall discuss the actions of  the automorphism $u_{K_*}$ on the eleven representations of $\Delta(27)$. Similar to the automorphisms $u_{S}$ and $u_{T}$, the consistency condition Eq.~\eqref{eq:con_uK} can be understood as a similarity transformation between the representations $\rho^*$ and $\rho\circ u_{K_*}$. The action of the outer automorphism $u_{K_*}$ on a irreducible representation of $\Delta(27)$ is defined as $u_{K_*}:\rho\rightarrow \rho\circ u_{K_*}$. The automorphism $u_{K_*}$ in Eq.~\eqref{eq:CPfromDelta27} acts on the eleven irreducible representations of $\Delta(27)$ as
\begin{eqnarray}
\nonumber u_{K_*}:&& \bm{1_{0,0}}\leftrightarrow \bm{1_{0,0}}, \qquad \bm{1_{0,1}}\leftrightarrow \bm{1_{0,2}}, \qquad \bm{1_{1,1}}\leftrightarrow\bm{1_{1,1}}, \qquad \bm{1_{2,2}}\leftrightarrow\bm{1_{2,2}}, \\
&& \bm{1_{1,0}}\leftrightarrow\bm{1_{1,2}}, \qquad  \bm{1_{2,0}} \leftrightarrow \bm{1_{2,1}}, \qquad \bm{3}\leftrightarrow \bm{\bar{3}}\,,
\end{eqnarray}
which are shown in table~\ref{tab:character_Delta27}. Therefore in order to implement the CP-like symmetry in the context of $\Delta(27)$ flavor symmetry, the fields in the representations $\bm{1_{1,1}}$ and $\bm{1_{2,2}}$ have to appear in pair, and the same holds true for the representations $\bm{1_{1,0}}$, $\bm{1_{2, 1}}$ as well as $\bm{1_{1, 2}}$, $\bm{1_{2,0}}$. Furthermore, considering the action of $u_S$, $u_T$  given in Eq.~\eqref{eq:uS_uT_singlets}, we find that the eight non-trivial singlet representations of $\Delta(27)$ can be classified into three categories: $\bm{1_{0,1}}\oplus\bm{1_{0,2}}=\bm {2_1}$, $\bm{1_{1,1}}\oplus\bm{1_{2,2}}=\bm{2_2}$ and $\bm{1_{1,0}}\oplus\bm{1_{2,0}}\oplus\bm{1_{1,2}}\oplus\bm{1_{2,1}}=\bm{2_3}\oplus\bm{2_4}$, as shown in figure~\ref{fig:action-uS-uT-uKstar}. Taking into account the interplay between CP and the $S_3$ modular group further, one find the action of $u_{K_*}$ on the irreducible representations of EFG $\Delta(27)\rtimes S_{3}$ as follow,
\begin{eqnarray}
\nonumber u_{K^{*}}&:& \bm{1^{a}_{0,0}}\leftrightarrow \bm{1^{a}_{0,0}}, \qquad \bm{2_{0}}\leftrightarrow \bm{2_{0}}, \qquad \bm{2_{1,0}}\leftrightarrow \bm{2_{1,0}}, \qquad \bm{2_{1,1}}\leftrightarrow \bm{2_{1,2}}, \qquad \bm{2_{2,m}}\leftrightarrow \bm{2_{2,m}},   \\
\label{eq:uKstar-rep}&& \bm{2_{3,m}}\leftrightarrow \bm{2_{4,m}}, \qquad \bm{3^{a}}\leftrightarrow \bm{\bar{3}^{a}}, \qquad \bm{6}\leftrightarrow\bm{\bar{6}}\,.
\end{eqnarray}
Therefore $u_{K^{*}}$ is not a class-inverting automorphism of the EFG $\Delta(27)\rtimes S_{3}$, the CP transformation corresponding to $u_{K^{*}}$ can be consistently imposed as a symmetry if a model contains only a subset of the irreducible representations $\bm{1^{a}_{0,0}}$, $\bm{2_{0}}$,  $\bm{2_{1,0}}$, $\bm{2_{2,m}}$, $\bm{3^{a}}$, $\bm{\bar{3}^{a}}$, $\bm{6}$, $\bm{\bar{6}}$ which are mapped to their complex conjugates by $u_{K^{*}}$~\cite{Chen:2014tpa,Ratz:2019zak}.
Solving the consistency conditions of Eqs.~(\ref{eq:con_uK}, \ref{eq:consistency_K*_mod}) for $g=A, B$, we find the expressions of the CP-like transformations are given by,
\begin{eqnarray}
\nonumber  &&\rho_{\bm{1^{a}_{0, 0}}}(K_*)=1,\quad \rho_{\bm{2_{0}}}(K_*)=\rho_{\bm{2_{1,0}}}(K_*)=\mathbb{1}_{2}
\,, \quad
\rho_{\bm{2_{2,m}}}(K_*)=\left(
\begin{array}{cc}
	0 & 1 \\
	1 & 0 \\
\end{array}
\right)\,, \\
\nonumber  && \rho_{\bm{3^{a}}}(K_*)=\rho^{*}_{\bm{\bar{3}^{a}}}(K_*)=\frac{1}{\sqrt{3}}\left(
\begin{array}{ccc}
	\omega ^2 & 1 & 1 \\
	1 & \omega ^2 & 1 \\
	1 & 1 & \omega ^2 \\
\end{array}
\right)\,,\\
\label{eq:gCP-transf}&&\rho_{\bm{6}}(K_*)=\rho^{*}_{\bm{\bar{6}}}(K_*)=\frac{1}{\sqrt{3}}\left(
	\begin{array}{ccc}
		\omega ^2\mathbb{1}_{2} & \mathbb{1}_{2} & \mathbb{1}_{2} \\
		\mathbb{1}_{2} & \omega ^2\mathbb{1}_{2} & \mathbb{1}_{2} \\
		\mathbb{1}_{2} & \mathbb{1}_{2} & \omega ^2\mathbb{1}_{2} \\
	\end{array}
	\right)\,,
\end{eqnarray}
where the overall phase is dropped.

We turn to discuss the CP transformation of modular form multiplets. As the representations of the finite modular group generators $S$ and $T$ are unitary and symmetric in our basis given in Eq.~\eqref{eq:Tp_irre}, the consistency condition of Eq.~\eqref{eq:consistency_K*_mod} fixes the CP-like transformation matrix $\rho({K}_*)$ be an identity matrix up to an overall phase, i.e.
\begin{equation}\label{eq:rho_K_Tp}
\rho_{\bm{r}}({K}_*)=\Id_{\bm{r}}, \qquad \bm{r}=\bm{1}, \,\bm{1}' ,\,\bm{2}\,.
\end{equation}
From the transformation property of modular multiplets in  Eq.~\eqref{eq:K*_to_matter} and the representation matrix of $\rho({K}_*)$ above,  one can obtain~\cite{Novichkov:2019sqv,Ding:2021iqp}
\begin{equation}\label{eq:Ytau_K_Tra}
Y^{(k_{Y})}_{\bm{r}}(\tau)~\stackrel{{K}_*}{\longrightarrow}~Y^{(k_{Y})}_{\bm{r}}(-\bar{\tau})=(Y^{(k_{Y})}_{\bm{r}}(\tau))^*\,,
\end{equation}
where $Y^{(k_{Y})}_{\bm{r}}(\tau)$ denotes any level 2 modular form multiplets at weight $k_Y$ in the irreducible representation $\bm{r}$ of the finite modular group $S_{3}$. The modular multiplets of level 2 up to weight 8 are shown in Appendix~\ref{sec:S3_group}.

\section{\label{sec:mod_general}K\"ahler potential and superpotential invariant under EFG $\Delta(27)\rtimes S_{3}$  }

From Eq.~\eqref{eq:uS_uT_singlets}, we find that any one of the eight nontrivial singlets of $\Delta(27)$ is mapped to another under the action of $S_{3}$ modular symmetry. If one field is assigned to be a nontrivial singlet of $\Delta(27)$, the EFG $\Delta(27)\rtimes S_{3}$ requires the presence of another field in the nontrivial singlet related by the outer automorphism $u_{S}$ and $u_{T}$. As a result, the eight nontrivial one-dimensional representations of $\Delta(27)$ have to be arranged into four doublet $\bm{2_{i}}$ ($i=1,2,3,4$) shown in Eq.~\eqref{eq:4-red-doublet-D27}. Notice that the two component fields of $\bm{2_{i}}$ should transform in the same way under both SM gauge group and the auxiliary cyclic group. Hence the three generations of quarks and leptons can be assigned to transform as triplet $\bm{3}$ (or $\bm{\bar{3}}$), one trivial singlet $\bm{1_{0,0}}$ plus one reducible doublet $\bm{2_{i}}$ or three trivial singlets $\bm{1_{0,0}}$ of $\Delta(27)$. The Higgs fields $H_u$ and $H_d$ are assumed to be invariant under $\Delta(27)$ if no additional Higgs are introduced. Nevertheless, flavons can be assigned to trivial singlet, triplet and reducible doublets $\bm{2_{i}}$  of the traditional flavor symmetry $\Delta(27)$. As shown in section~\ref{subsec:gCP}, the matter fields and flavons can only in the irreducible representations $\bm{1^{a}_{0,0}}$, $\bm{2_{0}}$,  $\bm{2_{1,0}}$, $\bm{2_{2,m}}$, $\bm{3^{a}}$, $\bm{\bar{3}^{a}}$, $\bm{6}$, $\bm{\bar{6}}$, if the CP-like symmetry corresponding to $u_{K^{*}}$ is imposed in a model with the EFG $\Delta(27)\rtimes S_{3}$.

In this section, we shall preform a general analysis for the K\"ahler potential $\mathcal{K}$ and the superpotential $\mathcal{W}$ which are invariant under the EFG $\Delta(27)\rtimes S_{3}$ in the framework of $\mathcal{N}=1$ global supersymmetry. In the approach of EFG, the general form of $\mathcal{K}$ and $\mathcal{W}$ depend on level 2 modular forms which can be arranged into multiplets of $S_{3}$. To obtain the general forms of K\"ahler potential and superpotential, we assume that the modular multiplets of level 2 and weight $k_{Y}$ comprise all possible irreducible multiplets of $S_{3}$, i.e.
\begin{equation}
 Y^{(k_{Y})}_{\bm{1}}(\tau)=Y_1\,,\qquad
Y^{(k_{Y})}_{\bm{1'}}(\tau)=Y_2\,,\qquad
Y^{(k_{Y})}_{\bm{2}}(\tau)=\begin{pmatrix}
	Y_3 \\
	Y_4
\end{pmatrix}\,.
\end{equation}
If some modular multiplets are absent for a given weight, the corresponding modular forms $Y_i$ must be set to zero. Notice that the contributions of the linearly independent modular form multiplets in the same representation of $S_3$ take a similar form. In the following, we give the most general form of the K\"ahler potential for different possible representation assignments of the three generation matter fields under $\Delta(27)$.

\subsection{\label{sec:Kahlerpotential} K\"ahler potential}

It is known that the  K\"ahler potential admits many terms compatible with modular symmetry and they could induce sizable corrections to the fermion masses and mixing parameters~\cite{Chen:2019ewa}. The EFG provides a scheme to control the K\"ahler potential through the interplay of modular symmetry and traditional flavor symmetry~\cite{Nilles:2020nnc,Nilles:2020kgo}. In the present work, K\"ahler potential is required to be invariant under the actions of the EFG $\Delta(27)\rtimes S_{3}$. The traditional flavor group $\Delta(27)$ can impose severe constraints on the K\"ahler potential and the corresponding higher order corrections are suppressed by powers of $\langle\Phi\rangle/\Lambda$, where $\langle\Phi\rangle$ and $\Lambda$ represent the VEVs of the flavon fields and the cutoff scale, respectively. If the three generations of matter fields $\psi$ transform as $\bm{1^{a}_{0,0}}$ of $\Delta(27)\rtimes S_{3}$, they are generally distinguished by the different charges under auxiliary Abelian symmetry, so that the K\"ahler metric would be diagonal and the kinetic terms can be changed to canonical form by field redefinition. Hence we shall concentrate on the assignments that the three generators matter fields $\psi$ transform as $\bm{3^a}$, $\bm{\bar{3}^{a}}$ or $\bm{1^{a}_{0,0}}\oplus\bm{2_{i,m}}$ under $\Delta(27)\rtimes S_{3}$ EFG in this section.

\subsubsection{\label{subsubsec:kahler1} K\"ahler potential for $\psi\sim \mathbf{3^{a}}$ or $\mathbf{\bar{3}^{a}}$}

From Eq.~\eqref{eq:3DS3fromD27matrices}, we see that the representations $\bm{3^{0}}$ and $\bm{3^{1}}$ are different in the overall sign of the modular generators $S$ and $T$. Hence the same K\"ahler potential invariant under EFG would be obtained for both assignments $\psi\sim\bm{3^{0}}$ and $\psi\sim\bm{3^{1}}$. Without loss of generality, we shall consider $\psi\sim\bm{3^{0}}$ in the following. Then the general form of the leading order (LO) K\"ahler potential is given by
\begin{equation}\label{eq:kahler_LO1}
\mathcal{K}_{\rm LO} = \sum_{k,\bm{r_1},\bm{r_2},s} (-i \tau+ i \bar \tau)^{-k_{\psi}+k}
\left( Y^{(k)\dagger}_{\bm{r_1}} Y^{(k)}_{\bm{r_2}}  \psi^{\dagger} \psi\right)_{\bm{1^{0}_{0,0}},s}\,,
\end{equation}
where we have to sum over the even weights $k\in2\mathbbm{N}$, the representation $\bm{r}$ of all linearly independent modular multiplets $Y^{(k)}_{\bm{r}}(\tau)$ and all $\Delta(27)\rtimes S_{3}$ singlet contractions labelled by the index $s$. We have omitted the coupling constant of each contractions in Eq.~\eqref{eq:kahler_LO1}, and the modular form of weight 0 is taken to be $Y^{(0)}_{\bm{r}}=1$. The terms of $k=0$ will give the minimal K\"ahler potential. The modulus $\tau$ as well as modular forms are invariant under the action of the flavor symmetry $\Delta(27)$, consequently invariance under $\Delta(27)$ requires $\psi^{\dagger}\psi$ should contract to a trivial singlet of $\Delta(27)$, i.e.
\begin{equation}\label{eq:D27_inv_cont}
\sum_{\bm{\mathcal{R}}}\left(\psi^{\dagger}\psi\right)_{\bm{\mathcal{R}}}=\left(\psi^{\dagger}\psi\right)_{\bm{1^{0}_{0,0}}}=\psi^{\dagger}_1\psi_1+\psi^{\dagger}_2\psi_2+\psi^{\dagger}_3\psi_3\,,
\end{equation}
where $\left(...\right)_{\bm{\mathcal{R}}}$ denotes $\Delta(27)$ invariant  contractions and the subscript $\bm{\mathcal{R}}$ refers to those EFG irreducible multiplets with generators $A$ and $B$ being unit matrices. Then $\bm{\mathcal{R}}$ can be the representations $\bm{1^{0}_{0,0}}$, $\bm{1^{1}_{0,0}}$ and $\bm{2_{0}}$. We will adopt this convention in the following. Because the combination $\psi^{\dagger}_1\psi_1+\psi^{\dagger}_2\psi_2+\psi^{\dagger}_3\psi_3=\left(\psi^{\dagger}\psi\right)_{\bm{1^{0}_{0,0}}}$ in Eq.~\eqref{eq:D27_inv_cont} is invariant under the finite modular group $S_{3}$, the contraction of modular forms $Y^{(k)\dagger}_{\bm{r_1}} Y^{(k)}_{\bm{r_2}}$ must be invariant under $S_{3}$ as well. From the Kronecker products of $S_{3}$ in Eq.~\eqref{eq:S3_KP}, we find that $\bm{r_1}$ and $\bm{r_2}$ should be the same representation $\bm{r_1}=\bm{r_2}=\bm{r}$ of $S_{3}$. Thus the EFG $\Delta(27)\rtimes S_{3}$ constrains the general form of the LO K\"ahler potential as follow,
\begin{equation}\label{eq:Kaehler_LO1_final}
\mathcal{K}_{\rm LO}=\left(\psi^{\dagger}_1\psi_1+\psi^{\dagger}_2\psi_2+\psi^{\dagger}_3\psi_3\right)\sum_{k,\bm{r}}  (-i \tau+ i \bar \tau)^{-k_{\psi}+k}
\left( Y^{(k)\dagger}_{\bm{r}} Y^{(k)}_{\bm{r}} \right)_{\bm{1^{0}_{0,0}}} \,.
\end{equation}
One can straightforwardly read off the K\"ahler metric which is proportional to a unit matrix, and the minimal K\"ahler potential is reproduced.
We needs to rescale the supermultiplets of the theory in order to get canonical kinetic terms. The effect of such rescaling can be compensated by redefining the couplings of the superpotential.

The next-to-leading-order (NLO) corrections to the K\"ahler potential contains a flavon $\Phi$, and it can be written as
\begin{equation}\label{eq:kahler_NLO1}
\mathcal{K}_{\rm NLO} =\frac{1}{\Lambda} \sum_{k_{1},\bm{r_1},\bm{r_2}, s} (-i \tau+ i \bar \tau)^{-k_{\psi}+k_{1}}
\left( Y^{(k_{1})\dagger}_{\bm{r_1}} Y^{(k_{1}+k_{\Phi})}_{\bm{r_2}}  \psi^{\dagger} \psi\Phi\right)_{\bm{1^{0}_{0,0}},s}+\text{h.c.}\,,
\end{equation}
where $k_{\Phi}$ is the modular weight  of the flavon $\Phi$. Comparing the expressions of $\mathcal{K}_{\rm LO}$ in Eq.~\eqref{eq:kahler_LO1} and $\mathcal{K}_{\rm NLO}$ in Eq.~\eqref{eq:kahler_NLO1}, we find that the flavon $\Phi$ can contribute to $\mathcal{K}_{\rm NLO}$ if and only if $\Phi$ is invariant under the auxiliary symmetry group. From the $\Delta(27)$ tensor product $\bm{3}\otimes\bm{\bar{3}} =\sum^{2}_{r,s=0} \bm{1_{r,s}}$ and the transformation properties of the nine contraction singlets of $\psi^{\dagger}\psi$, we find they can be arranged into one trivial singlet and four doublets of $\Delta(27)\rtimes S_{3}$, i.e.
\begin{eqnarray}
\nonumber && \left(\psi^{\dagger}\psi\right)_{\bm{1^0_{0,0}}}=\psi^{\dagger}_1\psi_1 +\psi^{\dagger}_2\psi_2+\psi^{\dagger}_3\psi_3\,, \\
\nonumber&& \left(\psi^{\dagger}\psi\right)_{\bm{2_{1,0}}}=\left(\begin{array}{c}
\psi^{\dagger}_1\psi_2 +\psi^{\dagger}_2\psi _3+\psi^{\dagger} _3\psi_1 \\
\psi^{\dagger}_1\psi_3+\psi^{\dagger}_2\psi_1+\psi^{\dagger}_3\psi_2
\end{array}\right)\,,\\
\nonumber	&& \left(\psi^{\dagger}\psi\right)_{ \bm{2_{2,2}}}=\left(\begin{array}{c}
\psi^{\dagger}_1\psi_2+\omega^2\psi^{\dagger}_2\psi_3+\omega\psi^{\dagger}_3\psi_1\\
\omega^2 \psi^{\dagger}_1\psi_3+\psi^{\dagger}_2\psi_1+\omega\psi^{\dagger}_3\psi_2
\end{array}\right)\,,\\
\nonumber	&& \left(\psi^{\dagger}\psi\right)_{ \bm{2_{3,2}}}=\left(\begin{array}{c}
\psi^{\dagger}_1 \psi_1+ \omega^2\psi^{\dagger}_2\psi_2 + \omega\psi^{\dagger}_3\psi_3 \\
\omega^2\psi^{\dagger}_1 \psi_1+  \psi^{\dagger}_2\psi_2 + \omega\psi^{\dagger}_3 \psi_3
\end{array}\right)\,,\\
\label{eq:LLb_contract}&& \left(\psi^{\dagger}\psi\right)_{ \bm{2_{4,2}}}=\left(\begin{array}{c}
\omega\psi^{\dagger}_1\psi_3+\psi^{\dagger}_2\psi_1+\omega^2\psi^{\dagger}_3\psi_2 \\
\psi^{\dagger}_1\psi_2+\omega \psi^{\dagger}_2\psi_3+\omega^2\psi^{\dagger}_3\psi_1
\end{array}\right)\,.
\end{eqnarray}
The contractions of $\psi^{\dagger}\psi\Phi$ should be invariant under $\Delta(27)$, so that the flavon $\Phi_{i}$ must transform as $\bm{1_{0,0}}$ or $\bm{2_{i}}$ under $\Delta(27)$. For $\Phi\sim\bm{1^a_{0,0}}$, we find
\begin{equation}
\left( Y^{(k_{1})\dagger}_{\bm{r_1}} Y^{(k_{1}+k_{\Phi})}_{\bm{r_2}}  \psi^{\dagger} \psi\Phi\right)_{\bm{1^0_{0,0}}}=\left( Y^{(k_{1})\dagger}_{\bm{r_1}} Y^{(k_{1}+k_{\Phi})}_{\bm{r_2}} \right)_{\bm{1^{[2-a]}_{0,0}}}\left( \psi^{\dagger} \psi\right)_{\bm{1^0_{0,0}}}\Phi\,,	
\end{equation}
where the notation $[2-a]$ is defined as $2-a$ modulo 2 and this contraction result is proportional to $\left( \psi^{\dagger} \psi\right)_{\bm{1^0_{0,0}}}=\psi^{\dagger}_1\psi_1+\psi^{\dagger}_2\psi_2+\psi^{\dagger}_3\psi_3$. It can be absorbed into the LO K\"ahler potential in Eq.~\eqref{eq:Kaehler_LO1_final}, and provides no correction to fermion masses and mixing parameters. If a model contains a flavon  $\Phi=(\phi_{1},\phi_{2})^{T}\sim\bm{2_{i,m}}$ which is chargeless under the auxiliary group, the NLO K\"ahler potential $\mathcal{K}_{\rm NLO}$ invariant under the EFG $\Delta(27)\rtimes S_{3}$ can be expanded as
\begin{eqnarray}
\nonumber \mathcal{K}_{\rm NLO} &=& \frac{1}{\Lambda}\sum_{k_{1},\bm{r_1},\bm{r_2},\bm{\mathcal{R}}}(-i \tau+ i \bar \tau)^{-k_{\psi}+k_{1}}
\left( Y^{(k_{1})\dagger}_{\bm{r_1}} Y^{(k_{1}+k_{\Phi})}_{\bm{r_2}} \right)_{\bm{\mathcal{R}}} \left(\left(\psi^{\dagger} \psi\right)_{ \bm{2_{1,0}}}\Phi\right)_{\bm{\mathcal{R}}}\\
\nonumber&&+\frac{1}{\Lambda}\sum_{k_{1},\bm{r_1},\bm{r_2},\bm{\mathcal{R}}}(-i \tau+ i \bar \tau)^{-k_{\psi}+k_{1}}
\left( Y^{(k_{1})\dagger}_{\bm{r_1}} Y^{(k_{1}+k_{\Phi})}_{\bm{r_2}} \right)_{\bm{\mathcal{R}}} \left(\left(\psi^{\dagger} \psi\right)_{\bm{2_{2,2}}}\Phi\right)_{\bm{\mathcal{R}}}\\
\nonumber&&+\frac{1}{\Lambda}\sum_{k_{1},\bm{r_1},\bm{r_2},\bm{\mathcal{R}}}(-i \tau+ i \bar \tau)^{-k_{\psi}+k_{1}}
\left( Y^{(k_{1})\dagger}_{\bm{r_1}} Y^{(k_{1}+k_{\Phi})}_{\bm{r_2}} \right)_{\bm{\mathcal{R}}} \left(\left(\psi^{\dagger} \psi\right)_{ \bm{2_{3,2}}}\Phi\right)_{\bm{\mathcal{R}}}\\
\label{eq:Kaehler_NLO}&&+\frac{1}{\Lambda}\sum_{k_{1},\bm{r_1},\bm{r_2},\bm{\mathcal{R}}}(-i \tau+ i \bar \tau)^{-k_{\psi}+k_{1}}
\left( Y^{(k_{1})\dagger}_{\bm{r_1}} Y^{(k_{1}+k_{\Phi})}_{\bm{r_2}} \right)_{\bm{\mathcal{R}}} \left(\left(\psi^{\dagger} \psi\right)_{ \bm{2_{4,2}}}\Phi\right)_{\bm{\mathcal{R}}}\,,
\end{eqnarray}
where the subscript $\bm{\mathcal{R}}$ is defined below Eq.~\eqref{eq:D27_inv_cont}. For each possible flavon $\Phi=(\phi_{1},\phi_{2})^{T}\sim\bm{2_{i,m}}$, only the contractions in the $i$th row in Eq.~\eqref{eq:Kaehler_NLO} are not vanishing and the non-vanishing contractions  $(\psi^{\dagger}\psi\Phi)_{\bm{\mathcal{R}}}$ can be obtained from Eq.~\eqref{eq:two_doublet_cont}. Then the explicit expression of $\mathcal{K}_{\rm NLO}$ can be written out. If the corrections from this $\mathcal{K}_{\rm NLO}$ are  considered, one can check that the corresponding K\"ahler metric is always not proportional to a unit matrix, and  the non-canonical corrections are suppressed by $\langle\Phi\rangle/\Lambda$ in comparison with $\mathcal{K}_{\rm LO}$.

If a model does not contain a flavon $\Phi$ which a doublet of $\Delta(27)$ and invariant under auxiliary cyclic symmetries, the corrections to fermion masses and mixings arise from the next-to-next-to-leading order (NNLO) terms of the K\"ahler potential. Without loss of generality, the most general form of the NNLO corrections to the K\"ahler potential can be written as
\begin{equation}\label{eq:kahler_NNLO1}
\mathcal{K}_{\rm NNLO} = \frac{1}{\Lambda^2}\sum_{k_{1}, \bm{r_1},\bm{r_2}, s} (-i \tau+ i \bar \tau)^{-k_{\psi}-k_{\Theta}+k_{1}}\left( Y^{(k_{1})\dagger}_{\bm{r_1}} Y^{(k_{1}+k_{\Phi}-k_{\Theta})}_{\bm{r_2}} \psi^{\dagger}\psi\Theta^{\dagger}\Phi\right)_{\bm{1^0_{0,0}},s}+\text{h.c.}\,,
\end{equation}
where each team involves two generic flavons $\Phi$ and $\Theta$ which could be the identical fields. Analogously the traditional flavor symmetry $\Delta(27)$ requires that $\psi^{\dagger} \psi\Theta^{\dagger}\Phi$ should be invariant under $\Delta(27)$, thus only the following contractions are allowed,
\begin{equation}\label{eq:psi-psiDag-theta-phi}
\left(\psi^{\dagger} \psi\Theta^{\dagger}\Phi\right)_{\bm{\mathcal{R}}}= \left(\psi^{\dagger}\psi\right)_{\bm{1^0_{0,0}}}\left(\Theta^{\dagger}\Phi\right)_{\bm{\mathcal{R}}}+\sum_{i=1}^{4}\left[\left(\psi^{\dagger}\psi\right)_{\bm{2_{i,m}}}\left(\Theta^{\dagger}\Phi\right)_{\bm{2_{i,n}}}\right]_{\bm{\mathcal{R}}}\,,
\end{equation}
which contract with the modular form $Y^{(k_{1})\dagger}_{\bm{r_1}} Y^{(k_{1}+k_{\Phi}-k_{\Theta})}_{\bm{r_2}}$ to form $S_{3}$ invariants. We see that $\Theta^{\dagger}\Phi$ should be the invariant singlet $\bm{1_{0,0}}$ or the reducible doublet $\bm{2_{i}}$ of the $\Delta(27)$ flavor symmetry. The flavons $\Theta$ and $\Phi$ can transform as $\bm{1_{0,0}}$, $\bm{2_{i}}$, $\bm{3}$ or $\bm{\bar{3}}$ under $\Delta(27)$. The expressions of $\left(\psi^{\dagger}\psi\right)_{\bm{1^0_{0,0}}}$ and $\left(\psi^{\dagger}\psi\right)_{\bm{2_{i,m}}}$ are given in Eq.~\eqref{eq:LLb_contract}. From Eq.~\eqref{eq:LLb_contract}, we find that the first term $\left(\psi^{\dagger}\psi\right)_{\bm{1^0_{0,0}}}\left(\Theta^{\dagger}\Phi\right)_{\bm{\mathcal{R}}}$ leads to a K\"ahler metric proportional to a unit matrix and its contribution can be absorbed by $\mathcal{K}_{\rm LO}$ while the second contraction $\left[\left(\psi^{\dagger}\psi\right)_{\bm{2_{i,m}}}\left(\Theta^{\dagger}\Phi\right)_{\bm{2_{i,n}}}\right]_{\bm{\mathcal{R}},s}$ can be obtained from Eq.~\eqref{eq:two_doublet_cont}:
\begin{eqnarray}
\nonumber \sum_{\bm{\mathcal{R}}}\left[\left(\psi^{\dagger}\psi\right)_{\bm{2_{i,m}}}\left(\Theta^{\dagger}\Phi\right)_{\bm{2_{i,n}}}\right]_{\bm{\mathcal{R}}} &=& \left[\left(\psi^{\dagger}\psi\right)_{\bm{2_{i,m}}}\left(\Theta^{\dagger}\Phi\right)_{\bm{2_{i,n}}}\right]_{\bm{1^{0}_{0,0}}} \\
\nonumber    &&+\left[\left(\psi^{\dagger}\psi\right)_{\bm{2_{i,m}}}\left(\Theta^{\dagger}\Phi\right)_{\bm{2_{i,n}}}\right]_{\bm{1^{1}_{0,0}}} \qquad \text{for} \quad m=n\,, \\
\sum_{\bm{\mathcal{R}}}\left[\left(\psi^{\dagger}\psi\right)_{\bm{2_{i,m}}}\left(\Theta^{\dagger}\Phi\right)_{\bm{2_{i,n}}}\right]_{\bm{\mathcal{R}}} &=& \left[\left(\psi^{\dagger}\psi\right)_{\bm{2_{i,m}}}\left(\Theta^{\dagger}\Phi\right)_{\bm{2_{i,n}}}\right]_{\bm{2_{0}}} \qquad \text{for} \quad m\neq n\,.
\end{eqnarray}
These terms will give rise to off-diagonal elements of the K\"ahler metric in a general model. Hence the K\"ahler potential $\mathcal{K}_{\rm NNLO}$ generally yield deviations from canonical kinetic terms of quark/lepton fields after the flavons develop VEVs, unless all the flavons $\Theta$  and $\Phi$ are invariant singlet of $\Delta(27)$. However, the induced corrections to the quark/lepton mixing parameters are suppressed by $\langle\Phi\rangle^2/\Lambda^2$ and they are negligible.
In the case of $\psi\sim\bm{\bar{3}^{a}}$, we reach the same results as those of $\psi\sim\bm{3^{a}}$ except that the K\"ahler metric becomes into the transpose.

\subsubsection{\label{subsubsec:kahler2} K\"ahler potential for $\psi\sim \mathbf{1^{a}_{0,0}}\oplus\mathbf{2_{i,m}}$}

The three generations of matter fields could be assigned to transform as reducible triplet of $\Delta(27)$. For instance, the first generation is an invariant singlet $\psi_{1}\sim\bm{1^{a}_{0,0}}$ under EFG and the other two generations form a doublet $\psi_{d}=(\psi_{2},\psi_{3})^T\sim\bm{2_{i,m}}$. At leading order, the K\"ahler potential for the matter fields $\psi$ can be written as
\begin{eqnarray}
\nonumber   \mathcal{K}_{\rm LO} &=& \sum_{k,\bm{r_1},\bm{r_2},s} (-i \tau+ i \bar \tau)^{-k_{\psi_1}+k}
\left( Y^{(k)\dagger}_{\bm{r_1}} Y^{(k)}_{\bm{r_2}}  \psi_{1}^{\dagger} \psi_{1}\right)_{\bm{1^{0}_{0,0}},s}
+(-i \tau+ i \bar \tau)^{-k_{\psi_d}+k}
\left( Y^{(k)\dagger}_{\bm{r_1}} Y^{(k)}_{\bm{r_2}}  \psi_{d}^{\dagger} \psi_{d}\right)_{\bm{1^{0}_{0,0}},s} \\
\nonumber &=&  \psi_{1}^{\dagger} \psi_{1} \sum_{k,\bm{r_1},\bm{r_2}}(-i \tau+ i \bar \tau)^{-k_{\psi_1}+k}
\left( Y^{(k)\dagger}_{\bm{r_1}} Y^{(k)}_{\bm{r_2}} \right)_{\bm{1^{0}_{0,0}}}\\
\nonumber &&+ \psi_{2}^{\dagger} \psi_{2} \sum_{k,\bm{r_1},\bm{r_2}}(-i \tau+ i \bar \tau)^{-k_{\psi_d}+k}\left[\left( Y^{(k)\dagger}_{\bm{r_1}} Y^{(k)}_{\bm{r_2}} \right)_{\bm{1^{0}_{0,0}}}+\left( Y^{(k)\dagger}_{\bm{r_1}} Y^{(k)}_{\bm{r_2}} \right)_{\bm{1^{1}_{0,0}}}  \right] \\
\label{eq:kahler_LO2} &&+\psi_{3}^{\dagger} \psi_{3} \sum_{k,\bm{r_1},\bm{r_2}}(-i \tau+ i \bar \tau)^{-k_{\psi_d}+k}\left[\left( Y^{(k)\dagger}_{\bm{r_1}} Y^{(k)}_{\bm{r_2}} \right)_{\bm{1^{0}_{0,0}}}-\left( Y^{(k)\dagger}_{\bm{r_1}} Y^{(k)}_{\bm{r_2}} \right)_{\bm{1^{1}_{0,0}}}  \right] \,.
\end{eqnarray}
Hence the resulting K\"ahler metric is diagonal while the diagonal entries are all different. When transform to the basis with canonical kinetic terms, we have to rescale the matter fields $\psi_{1,2,3}$. The effect of rescaling on the second and third generator fermion masses can not be absorbed into the parameters of superpotential. As a consequence, if the matter fields $\psi$ are assigned to be one trivial singlet $\bm{1_{0,0}}$ plus one reducible doublet $\bm{2_{i}}$ of $\Delta(27)$, the EFG $\Delta(27)\rtimes S_{3}$ does not efficiently restrict the K\"ahler potential so that the predictive power of modular symmetry would be reduced. Hence we shall not consider these reducible assignments for matter fields in the EFG model construction.

\subsection{Superpotential for fermion masses }

As has been shown in previous section, the corrections from K\"ahler potential to the fermion masses and flavor mixing are under control by the EFG $\Delta(27)\rtimes S_{3}$, if the matter fields are assigned to transform as trivial singlet or triplet under $\Delta(27)$. The EFG $\Delta(27)\rtimes S_{3}$ would play less rule for the invariant singlet assignment of the left-handed (LH) matter fields $\psi$ and right-handed (RH) matter fields $\psi^c$. Hence we will analyze the assignments that both $\psi$ and $\psi^c$ are $\Delta(27)$ triplet $\bm{3}$ or $\bm{\bar{3}}$, their modular weights are denoted as $k_{\psi}$ and $k_{\psi^c}$ respectively. In the paradigm of EFG, flavon fields are usually necessary to break the traditional flavor symmetry. For the concerned EFG $\Delta(27)\rtimes S_{3}$, the flavon $\Phi$ with modular weight $k_{\Phi}$ can transform as singlet $\bm{1_{0,0}}$, doublets $\bm{2_{i}}$ or triplets $\bm{3}$ and $\bm{\bar{3}}$ under $\Delta(27)$. Next we study the mass matrix of $\psi$ ($\psi^c$) for each possible representation assignments of the matter fields and flavon.

\begin{description}[labelindent=-0.7em, leftmargin=0.9em]
	
\item[~~(\lowercase\expandafter{\romannumeral1})] { $\psi, ~\psi^c,~ \Phi\sim \bm{3^a}$ or $ \bm{\bar{3}^a}$ }

In the first case, both $\psi$ and $\psi^c$ are triplets $\bm{3}$ (or $\bm{\bar{3}}$) of $\Delta(27)$. Invariance under the $\Delta(27)$ flavor symmetry entails the introduction of a triplet flavon $\Phi=(\phi_{1},\phi_{2},\phi_{3})^T$ which transforms as $\bm{3}$ or $\bm{\bar{3}}$ under $\Delta(27)$. For illustration, we proceed to analyze the superpotential for the representation assignment with $\psi\equiv\left(\psi_1, \psi_2, \psi_3\right)^T\sim \bm{3^0}$, $\psi^{c}\equiv\left(\psi^c_1, \psi^c_2, \psi^c_3\right)^{T}\sim \bm{3^0}$, $\Phi\equiv(\phi_1, \phi_2, \phi_3)^{T}\sim \bm{3^0}$, the superpotential of other assignments can be discussed analogously. Then the charged lepton/quark mass terms invariant under $\Delta(27)\times S_{3}$ can be generally written as
\begin{equation}
\mathcal{W}_{D}=\frac{1}{\Lambda}\sum_{\bm{r},s}c_{\bm{r},s}\left(Y^{(k_Y)}_{\bm{r}}\Phi \psi^{c}\psi\right)_{\bm{1^{0}_{0,0}},s}H_{u/d}\,,
\end{equation}
where one must sum over all modular multiplets of weight $k_Y$ and all independent singlet contractions labelled by the index $s$, and the modular weight $k_Y$ should fulfill $k_Y=k_{\psi}+k_{\psi^{c}}+k_{\Phi}$. It is convenient to firstly consider the constraints of the traditional flavor group $\Delta(27)$,  the general form of the superpotential $\mathcal{W}_{D}$ compatible with $\Delta(27)$ is given by
\begin{equation}\label{eq:WD1}
\mathcal{W}_{D}=\frac{1}{\Lambda}\sum_{\bm{r}} \Big(Y^{(k_Y)}_{\bm{r}}\big(c_{\bm{r},1}\mathcal{O}_{1}+c_{\bm{r},2}\mathcal{O}_{2}+c_{\bm{r},3}\mathcal{O}_{3}\Big)\Big)_{\bm{1^{0}_{0,0}}}H_{u,d}\,,
\end{equation}
where $\mathcal{O}_1$, $\mathcal{O}_2$ and $\mathcal{O}_3$ are three $\Delta(27)$ invariant contractions:
\begin{eqnarray}
\nonumber\mathcal{O}_1&=&\left(\Phi\left(\psi^c\psi\right)_{\bm{\overline{3}_{S_1}}}\right)_{\bm{1}_{0,0}}=\psi^c_1 \psi_{1} \phi_{1}+\psi^c_2 \psi_{2} \phi_{2}+\psi^c_3 \psi_{3} \phi_{3}\,,\\
\nonumber\mathcal{O}_2&=&\left(\Phi\left(\psi^c\psi\right)_{\bm{\overline{3}_{S_2}}}\right)_{\bm{1}_{0,0}}=\psi^c_1 \psi_{2} \phi_{3}+\psi^c_1 \psi_{3} \phi_{2}+\psi^c_2 \psi_{3} \phi_{1}+\psi^c_2 \psi_{1} \phi_{3}+\psi^c_3 \psi_{1} \phi_{2}+\psi^c_3 \psi_{2} \phi_{1}\,,\\
\mathcal{O}_3&=&\left(\Phi\left(\psi^c\psi\right)_{\bm{\overline{3}_{A}}}\right)_{\bm{1}_{0,0}}=\psi^c_1 \psi_{2} \phi_{3}-\psi^c_1 \psi_{3} \phi_{2}+\psi^c_2 \psi_{3} \phi_{1}-\psi^c_2 \psi_{1} \phi_{3}+\psi^c_3 \psi_{1} \phi_{2}-\psi^c_3 \psi_{2} \phi_{1}\,.~~~~
\end{eqnarray}
The modular transformation of the fields $\psi$, $\psi^{c}$, $\Phi$ is given in Eq.~\eqref{eq:3DS3fromD27matrices}, then we can obtain the actions of the modular generators $S$ and $T$ on the above combinations $\mathcal{O}_{1,2,3}$ as follows
\begin{eqnarray}
\nonumber&&\mathcal{O}_1\stackrel{S}{\longrightarrow}\mathcal{O}_1,\qquad \mathcal{O}_2\stackrel{S}{\longrightarrow}\mathcal{O}_2, \qquad \mathcal{O}_3\stackrel{S}{\longrightarrow}-\mathcal{O}_3\,,\\
&& \mathcal{O}_1\stackrel{T}{\longrightarrow}\mathcal{O}_1, \qquad
\mathcal{O}_2\stackrel{T}{\longrightarrow}\mathcal{O}_2, \qquad \mathcal{O}_3\stackrel{T}{\longrightarrow}-\mathcal{O}_3\,.
\end{eqnarray}
This implies that both $\mathcal{O}_1$ and $\mathcal{O}_2$ are invariant under $S_{3}$, while the combination $\mathcal{O}_3$ transforms as $\bm{1^{\prime}}$ under $S_{3}$.
Therefore only the modular form multiplets $Y^{(k_Y)}_{\bm{1}}=Y_1$ and $Y^{(k_Y)}_{\bm{1'}}=Y_2$ can contract with $\mathcal{O}_{1,2}$ and $\mathcal{O}_{3}$ respectively to produce mass terms invariant under $S_{3}$. Thus the most general superpotential which is invariant under the EFG $\Delta(27)\rtimes S_{3}$ is determined to be
\begin{equation}
\mathcal{W}_{D}=\frac{1}{\Lambda}\Big( \alpha_{1} Y_{1} \mathcal{O}_1+\alpha_{2} Y_{1}\mathcal{O}_2+\alpha_{3} Y_{2} \mathcal{O}_3\Big)H_{u,d}\,,
\end{equation}
with $\alpha_{1}\equiv c_{\bm{1},1}$, $\alpha_{2}\equiv c_{\bm{1}, 2}$ and $\alpha_{3}\equiv c_{\bm{1'}, 3}$. Note  the first two terms  arise from the coupling with $Y^{(k_Y)}_{\bm{1}}$ and the third term proportional to $\alpha_3$ arises from the coupling with $Y^{(k_Y)}_{\bm{1^\prime}}$. If $Y^{(k_Y)}_{\bm{1^\prime}}$ is absent in certain modular weight $k_{Y}$,  then $Y_{2}=0$ and the third term is vanishing. Therefore we get the following fermion mass matrix in the right-left
basis of $\psi^c M_{\psi}\psi$,
\begin{equation}
\label{eq:Ch_mass1} M_{\psi}= \frac{v_{u,d}}{\Lambda}\left[\alpha_{1}Y_{1}\left(
\begin{array}{ccc}
		\phi_{1} & 0 & 0 \\
		0 & \phi_{2} & 0 \\
		0 & 0 & \phi_{3} \\
\end{array}
	\right)+\alpha_{2}Y_{1}\left(
	\begin{array}{ccc}
		0 & \phi_{3} & \phi_{2} \\
		\phi_{3} & 0 & \phi_{1} \\
		\phi_{2} & \phi_{1} & 0 \\
	\end{array}
	\right)+\alpha_{3}Y_{2}\left(
	\begin{array}{ccc}
		0 & -\phi_{3} & \phi_{2} \\
		\phi_{3} & 0 & -\phi_{1} \\
		-\phi_{2} & \phi_{1} & 0 \\
	\end{array}
	\right)\right]\,,
\end{equation}
where $v_{u,d}=\langle H_{u,d}\rangle$ denote the VEVs of Higgs fields. If one does not impose the CP-like symmetry, all couplings $\alpha_{1,2,3}$ are generally complex so that the modular forms $Y_{1}$ and $Y_{2}$ in the mass matrix of Eq.~\eqref{eq:Ch_mass1} can be absorbed into the coupling constants. On the other hand, if the CP-like symmetry is included, the modular forms can not be absorbed by the coupling constants. With the CP transformation matrices in Eq.~\eqref{eq:gCP-transf}, we find that the CP invariance leads to the following constraints on the couplings $\alpha_1$, $\alpha_2$ and $\alpha_{3}$:
\begin{equation}\label{eq:CP_cons_Mpsi}
\alpha_{2}=\frac{\omega}{2}(\sqrt{3}\alpha^*_{1}-\alpha_{1})\,, \qquad \text{Arg}(\alpha_{3})=-\frac{\pi}{12}\,\,\,(\text{mod} \,\pi)\,.
\end{equation}
Consequently $\alpha_{1}$, $\alpha_{2}$ and $\alpha_{3}$ can be parameterized as
\begin{equation}
\alpha_{1}=\alpha_{r}+i\alpha_{i}, \qquad \alpha_{2}=\frac{\omega}{2}\left[(\sqrt{3}-1)\alpha_{r}-i(1+\sqrt{3})\alpha_{i}\right], \qquad \alpha_{3}=e^{-i\frac{\pi}{12}}\alpha^\prime_{3}\,,
\end{equation}
where $\alpha_{r}$, $\alpha_{i}$ and $\alpha^\prime_{3}$ are real. Thus the mass matrix $M_{\psi}$ including the phases of couplings explicitly is given by
\begin{eqnarray}
\nonumber 	M_{\psi}&=&\frac{v_{u,d}}{\Lambda}\left\{\frac{Y_{1}}{2}\left[\alpha_{r}\left(
\begin{array}{ccc}
		2 \phi_{1} & \left(\sqrt{3}-1\right) \omega  \phi_{3} & \left(\sqrt{3}-1\right) \omega  \phi_{2} \\
		\left(\sqrt{3}-1\right) \omega  \phi_{3} & 2 \phi_{2} & \left(\sqrt{3}-1\right) \omega  \phi_{1} \\
		\left(\sqrt{3}-1\right) \omega  \phi_{2} & \left(\sqrt{3}-1\right) \omega  \phi_{1} & 2 \phi_{3} \\
\end{array}\right)\right.\right. \\
\nonumber 	&&~~~\left.+i\alpha_{i}\left(
\begin{array}{ccc}
		2 \phi_{1} &-\left(\sqrt{3}+1\right)\omega  \phi_{3} & -\left(\sqrt{3}+1\right)\omega \phi_{2} \\
		-	\left(\sqrt{3}+1\right)\omega \phi_{3} & 2 \phi_{2} &- \left(\sqrt{3}+1\right)\omega  \phi_{1} \\
		-\left(\sqrt{3}+1\right) \omega \phi_{2} &- \left(\sqrt{3}+1\right) \omega \phi_{1} & 2 \phi_{3} \\
\end{array}\right)\right]\\
&&~~~\left.+e^{-i\frac{\pi}{12}}\alpha^\prime_{3}Y_{2}\left(
\begin{array}{ccc}
		0 & -\phi_{3} & \phi_{2} \\
		\phi_{3} & 0 & -\phi_{1} \\
		-\phi_{2} & \phi_{1} & 0 \\
\end{array}
\right)\right\}\,.
\end{eqnarray}
For the other three assignments of $\psi$, $\psi^c$ and $\Phi$ which are all triplets of $\Delta(27)$, we can analogously obtain the corresponding fermion mass matrices shown in table~\ref{tab:CL_mms}, and they are named as $W_{\psi2}$, $W_{\psi3}$ and $W_{\psi4}$ respectively. Note that the modular forms $Y_{1}$ and $Y_{2}$ in the mass matrices of $W_{\psi1}\sim W_{\psi4}$ can be absorbed by the coupling constants for models without CP-like symmetry, hence we essentially get the same fermion mass matrix and the modulus $\tau$ plays no role. If CP-like symmetry is imposed, the phases of couplings would be fixed and then the effect of the modular forms $Y_{1}$ and $Y_{2}$ can not be ignored. For the cases $W_{\psi3}$ and $W_{\psi4}$ in which $\psi$, $\psi^c$ and $\Phi$ transform as triplet $\bm{\bar{3}}$ under $\Delta(27)$, the CP symmetry enforces the couplings $\alpha_{1,2,3}$ to satisfy
\begin{equation}\label{eq:CP_cons_Mpsi2}
\alpha_{2}=\frac{\omega^2}{2}(\sqrt{3}\,\alpha^*_{1}-\alpha_{1})\,, \qquad \text{Arg}(\alpha_{3})=\frac{\pi}{12}\,\,\,(\text{mod} \,\pi)\,.
\end{equation}

\begin{table}[hptb!]
	\small
	\renewcommand{\arraystretch}{1.1}
\centering
\renewcommand{\tabcolsep}{1.1mm}
\begin{tabular}{|c|c|c|c|c|c|c|c|c|c|c|c|}  \hline\hline

$\mathcal{W}$ & $\Delta(27)\rtimes S_{3}$  & $Y^{(k_{Y})}_{\bm{r}}$ & $M_{\psi}$ & Constraints of CP symmetry \\  \hline

$\mathcal{W}_{\psi1}$ & $\psi,\,\psi^{c}\sim\bm{3^{0}}$, $\Phi\sim\bm{3^{0}}$   & $Y^{(k_{Y})}_{\bm{1}}$, $Y^{(k_{Y})}_{\bm{1^{\prime}}}$  & $M_{\psi}$ in Eq.~\eqref{eq:Ch_mass1} & Eq.~\eqref{eq:CP_cons_Mpsi} \\  \hline
		
$\mathcal{W}_{\psi2}$ & $\psi,\,\psi^{c}\sim\bm{3^{0}}$, $\Phi\sim\bm{3^{1}}$   &  $Y^{(k_{Y})}_{\bm{1}}$, $Y^{(k_{Y})}_{\bm{1^{\prime}}}$ & $M_{\psi}\left(Y_1\to Y_2,Y_{2}\to Y_{1}\right)$ & Eq.~\eqref{eq:CP_cons_Mpsi}  \\  \hline
		
$\mathcal{W}_{\psi3}$ & $\psi,\,\psi^{c}\sim\bm{\bar{3}^{0}}$, $\Phi\sim\bm{\bar{3}^{0}}$   &  $Y^{(k_{Y})}_{\bm{1}}$, $Y^{(k_{Y})}_{\bm{1^{\prime}}}$ & $M_{\psi}$ & Eq.~\eqref{eq:CP_cons_Mpsi2} \\  \hline
		
$\mathcal{W}_{\psi4}$ & $\psi,\,\psi^{c}\sim\bm{\bar{3}^{0}}$, $\Phi\sim\bm{\bar{3}^{1}}$   &  $Y^{(k_{Y})}_{\bm{1}}$, $Y^{(k_{Y})}_{\bm{1^{\prime}}}$ & $M_{\psi}\left(Y_1\to Y_2,Y_{2}\to Y_{1}\right)$ & Eq.~\eqref{eq:CP_cons_Mpsi2} \\  \hline
		
\multirow{2}{*}{$\mathcal{W}_{\psi5}$} & $\psi\sim\bm{3^{0}}$, $\psi^{c}\sim\bm{\bar{3}^{0}}$,    &  \multirow{2}{*}{$Y^{(k_{Y})}_{\bm{1}}$, $Y^{(k_{Y})}_{\bm{1^{\prime}}}$}  &   \multirow{2}{*}{$M^\prime_{\psi}$ in Eq.~\eqref{eq:Ch_mass2} } &    \multirow{2}{*}{ Eq.~\eqref{eq:CP_cons_Mpsip}}  \\
&  $\Phi\sim\bm{2_{1,0}}$  &  &   &   \\ \hline
\multirow{2}{*}{$\mathcal{W}_{\psi6}$} & $\psi\sim\bm{3^{0}}$, $\psi^{c}\sim\bm{\bar{3}^{0}}$,    &  \multirow{2}{*}{$Y^{(k_{Y})}_{\bm{1}}$, $Y^{(k_{Y})}_{\bm{1^{\prime}}}$}  &   \multirow{2}{*}{$D_{L}M^\prime_{\psi}\left(Y_1\to \omega^2Y_1,Y_{2}\to i\omega Y_{2}\right)D_{R}$ } &    \multirow{2}{*}{ Eq.~\eqref{eq:CP_cons_Mpsip}}  \\
&  $\Phi\sim\bm{2_{2,2}}$  &  &   &  \\ \hline

\multirow{2}{*}{$\mathcal{W}_{\psi7}$} & $\psi\sim\bm{3^{0}}$, $\psi^{c}\sim\bm{\bar{3}^{0}}$,    &  \multirow{2}{*}{$Y^{(k_{Y})}_{\bm{1}}$, $Y^{(k_{Y})}_{\bm{1^{\prime}}}$}  &  $\phi_{1}(\alpha_{1}Y_{1}+\alpha_{2}Y_{2})\text{diag}(\omega,\omega^2,1)$  &    \multirow{2}{*}{ ---}  \\
&  $\Phi\sim\bm{2_{3,2}}$  & & $+\phi_{2}(\alpha_{1}Y_{1}-\alpha_{2}Y_{2})\text{diag}(\omega,1,\omega^2)$ &    \\ \hline
		
\multirow{2}{*}{$\mathcal{W}_{\psi8}$} & $\psi\sim\bm{3^{0}}$, $\psi^{c}\sim\bm{\bar{3}^{0}}$,    &  \multirow{2}{*}{$Y^{(k_{Y})}_{\bm{1}}$, $Y^{(k_{Y})}_{\bm{1^{\prime}}}$}  &   \multirow{2}{*}{$P_{23}D_{R}M^\prime_{\psi}(P_{23}D_{R})^{\dagger}$ } &    \multirow{2}{*}{ ---}  \\
&  $\Phi\sim\bm{2_{4,2}}$  &  &   &  \\ \hline
		
\multirow{2}{*}{$\mathcal{W}_{\psi9}$} & $\psi\sim\bm{3^{0}}$, $\psi^{c}\sim\bm{\bar{3}^{0}}$,    &  \multirow{2}{*}{$Y^{(k_{Y})}_{\bm{2}}$}  &   \multirow{2}{*}{$M^{\prime\prime}_{\psi}$ in Eq.~\eqref{eq:Ch_mass3} } &    \multirow{2}{*}{ ---}  \\
&  $\Phi\sim\bm{2_{1,1}}$  &  &   &  \\ \hline

\multirow{2}{*}{$\mathcal{W}_{\psi10}$} & $\psi\sim\bm{3^{0}}$, $\psi^{c}\sim\bm{\bar{3}^{0}}$,    &  \multirow{2}{*}{$Y^{(k_{Y})}_{\bm{2}}$}  &   \multirow{2}{*}{$M^{\prime\prime}_{\phi}\left(\phi_{1}\to \omega^2\phi_{1},Y_{4}\to -Y_{4}\right)$ } &    \multirow{2}{*}{ ---}  \\
&  $\Phi\sim\bm{2_{1,2}}$  &  &   &  \\ \hline
		
\multirow{2}{*}{$\mathcal{W}_{\psi11}$} & $\psi\sim\bm{3^{0}}$, $\psi^{c}\sim\bm{\bar{3}^{0}}$,    &  \multirow{2}{*}{$Y^{(k_{Y})}_{\bm{2}}$}  &   \multirow{2}{*}{$\omega^2D_{L}M^{\prime\prime}_{\psi}D_{R}$  } &    \multirow{2}{*}{ $\Im\alpha=0$}  \\
&  $\Phi\sim\bm{2_{2,0}}$  &  &   &   \\ \hline
		
\multirow{2}{*}{$\mathcal{W}_{\psi12}$} & $\psi\sim\bm{3^{0}}$, $\psi^{c}\sim\bm{\bar{3}^{0}}$,    &  \multirow{2}{*}{$Y^{(k_{Y})}_{\bm{2}}$}  &   \multirow{2}{*}{$D_{L}M^{\prime\prime}_{\phi}\left(\phi_{1}\to \phi_{1},\phi_{2}\to \omega\phi_{2},Y_{4}\to -Y_{4}\right)D_{R}$} &    \multirow{2}{*}{ $\Im\alpha=0$}  \\
&  $\Phi\sim\bm{2_{2,1}}$  &  &   &  \\ \hline
		
\multirow{2}{*}{$\mathcal{W}_{\psi13}$} & $\psi\sim\bm{3^{0}}$, $\psi^{c}\sim\bm{\bar{3}^{0}}$,    &  \multirow{2}{*}{$Y^{(k_{Y})}_{\bm{2}}$}  &  $\alpha[ \phi_{1}(Y_{3}-iY_{4})\text{diag}(\omega,\omega^2,1)$ &    \multirow{2}{*}{ ---}  \\
&  $\Phi\sim\bm{2_{3,0}}$  &  &  $+\phi_{2}(Y_{3}+iY_{4})\text{diag}(\omega^2,\omega,1)]$ &   \\ \hline
		
\multirow{2}{*}{$\mathcal{W}_{\psi14}$} & $\psi\sim\bm{3^{0}}$, $\psi^{c}\sim\bm{\bar{3}^{0}}$,    &  \multirow{2}{*}{$Y^{(k_{Y})}_{\bm{2}}$}  &  $\alpha[ \phi_{1}(Y_{3}+iY_{4})\text{diag}(1,\omega,\omega^2)$ &    \multirow{2}{*}{ ---}  \\
&  $\Phi\sim\bm{2_{3,1}}$  &  &  $+\phi_{2}(Y_{3}-iY_{4})\text{diag}(\omega^2,\omega,1)]$ &   \\ \hline
		
\multirow{2}{*}{$\mathcal{W}_{\psi15}$} & $\psi\sim\bm{3^{0}}$, $\psi^{c}\sim\bm{\bar{3}^{0}}$,    &  \multirow{2}{*}{$Y^{(k_{Y})}_{\bm{2}}$}  &   \multirow{2}{*}{$P_{23}D^\dagger_{L}M^{\prime\prime}_{\psi}D^\dagger_{R}P_{23}$  } &    \multirow{2}{*}{ ---}  \\
&  $\Phi\sim\bm{2_{4,0}}$  &  &   &   \\ \hline
		
\multirow{2}{*}{$\mathcal{W}_{\psi16}$} & $\psi\sim\bm{3^{0}}$, $\psi^{c}\sim\bm{\bar{3}^{0}}$,    &  \multirow{2}{*}{$Y^{(k_{Y})}_{\bm{2}}$}  &   \multirow{2}{*}{$P_{23}D^\dagger_{L}M^{\prime\prime}_{\psi}\left(\phi_{1}\to \omega^2\phi_{1},Y_{4}\to -Y_{4}\right)D^\dagger_{R}P_{23}$  } &    \multirow{2}{*}{ ---}  \\
&  $\Phi\sim\bm{2_{4,1}}$  &  &   &  \\ \hline \hline
	
\end{tabular}
\caption{\label{tab:CL_mms} The fermion mass matrices for different representation assignments of matter fields $\psi$, $\psi^{c}$ and flavon $\Phi$, where $D_{L}=\text{diag}(1,\omega^2,\omega^2)$,  $D_{R}=\text{diag}(\omega^2,1,1)$ and $P_{23}$ is the permutation matrix interchanging the 2nd and 3rd rows of the $3\times3$ identity matrix. The last column gives the constraints of CP-like symmetry on the coupling constants. }
\end{table}

\item[~~(\lowercase\expandafter{\romannumeral2})]  $\psi \sim \bm{3^a} \,(\bm{\bar{3}^a})$, $\psi^c\sim \bm{\bar{3}^b}\,(\bm{3^b})$, $ \Phi\sim \bm{2_{i,m}} $ for $\rho_{\bm{2_{i,m}}}(A)=\rho_{\bm{2_{i,m}}}(ST)$

For the second case, the matter fields $\psi$ and $\psi^c$ are assigned to triplets $\bm{3}\,(\bm{\bar{3}})$ and $\bm{\bar{3}} (\bm{3})$ of $\Delta(27)$, respectively. From the Kronecker product $\bm{3}\otimes \bm{\bar{3}}=\sum^{2}_{r,s=0} \bm{1_{r,s}}$ in Eq.~\eqref{eq:KP_Delta27}, one see that the flavon could be absent, it should transform as $\bm{1_{0,0}}$ or reducible doublet $\bm{2_{i}}$ under $\Delta(27)$ if it is present in a model. If there is no flavon or the flavon is invariant under $\Delta(27)$, the mass matrix of the fermion $\psi$ is proportional to a unit matrix. For the case of doublet flavon $\Phi=(\phi_{1},\phi_{2})^T$ transforming as $\bm{2_{i,m}}$ in which the representation matrices of $A$ and modular transformation $ST$ are identical, the corresponding values of the indices $(i,m)$ can be found in Eq.~\eqref{eq:ST-eq-A}. As an example, we analyze the superpotential and the corresponding mass matrix for the assignment of $\psi\sim \bm{3^0}$, $\psi^c\sim \bm{\bar{3}^0}$ and $\Phi\sim\bm{2_{1,0}}$. Invariance under the action of $\Delta(27)$ requires the superpotential $\mathcal{W}_{D}$ should be of the following form
\begin{equation}\label{eq:D27_inv_SP_M2}
\mathcal{W}_{D}=\frac{1}{\Lambda}\sum_{\bm{r}} \Big(Y^{(k_Y)}_{\bm{r}}\big[c_{\bm{r},1}\mathcal{O}_{4}+c_{\bm{r},2} \mathcal{O}_{5}\big]\Big)_{\bm{1^{0}_{0,0}}}H_{u/d}\,,
\end{equation}
where the two $\Delta(27)$ invariant combinations $\mathcal{O}_4$ and $\mathcal{O}_5$ take the following form
\begin{equation}\label{eq:def_O4O5}
\mathcal{O}_4=\phi_{1} (\psi^{c}_{1} \psi_{3}+\psi^{c}_{2} \psi_{1}+\psi^{c}_{3} \psi_{2})\,,\qquad \mathcal{O}_5=\phi_{2} (\psi^{c}_{1} \psi_{2}+\psi^{c}_{2} \psi_{3}+\psi^{c}_{3} \psi_{1})\,.
\end{equation}
From the transformation properties of $\psi$, $\psi^{c}$ and $\Phi$ under the actions of $S_{3}$, we find that the modular transformations of $\mathcal{O}_{4,5}$ under $S$ and $T$ are given by
\begin{equation}
\mathcal{O}_4\stackrel{S}{\longrightarrow}\mathcal{O}_5,\qquad \mathcal{O}_4\stackrel{T}{\longrightarrow}\mathcal{O}_5\,,\qquad
\mathcal{O}_5\stackrel{S}{\longrightarrow}\mathcal{O}_4,
	\qquad \mathcal{O}_5\stackrel{T}{\longrightarrow}\mathcal{O}_4\,.
\end{equation}
Therefore $\mathcal{O}_4$ and $\mathcal{O}_5$ can be arranged into two singlets $\bm{1}$ and $\bm{1^\prime}$ of $S_{3}$ modular symmetry,
\begin{equation}
\mathcal{O}^\prime_4=\mathcal{O}_4+\mathcal{O}_5 ~~\sim ~~\bm{1}, \qquad \qquad  \mathcal{O}^\prime_5=\mathcal{O}_4-\mathcal{O}_5 ~~\sim ~~\bm{1^\prime}\,.
\end{equation}
Consequently only the singlet modular forms $Y^{(k_{Y})}_{\bm{1}}(\tau)$ and $Y^{(k_{Y})}_{\bm{1'}}(\tau)$ can combine with $\mathcal{O}^\prime_4$ and $\mathcal{O}^\prime_5$ respectively to form EFG invariant superpotential,
\begin{equation}
\mathcal{W}_{D}=\frac{1}{\Lambda}\Big( \alpha_{1} Y_{1}\mathcal{O}^\prime_4 + \alpha_{2} Y_{2} \mathcal{O}^\prime_5\Big)H_{u/d}\,,
\end{equation}
which leads to the following mass matrix,
\begin{eqnarray}
\nonumber M^{\prime}_{\psi}&=& \frac{v_{u,d}}{\Lambda}\left[\alpha_{1}Y_{1}\left(
\begin{array}{ccc}
0 & \phi_{2} & \phi_{1} \\
\phi_{1} & 0 & \phi_{2} \\
\phi_{2} & \phi_{1} & 0
\end{array}\right)+\alpha_{2}Y_{2}\left(
\begin{array}{ccc}
0 & -\phi_{2} & \phi_{1} \\
\phi_{1} & 0 & -\phi_{2} \\
-\phi_{2} & \phi_{1} & 0
\end{array}
\right)\right] \\
\label{eq:Ch_mass2} &=&\frac{v_{u,d}}{\Lambda}\left[\alpha^{\prime}_{1}\phi_{1}\left(
\begin{array}{ccc}
0 & 0 & 1 \\
1 & 0 & 0 \\
0 & 1 & 0
\end{array}\right)+\alpha^{\prime}_{2}\phi_{2}\left(
\begin{array}{ccc}
0 & 1 & 0 \\
0 & 0 & 1 \\
1 & 0 & 0
\end{array}
\right)\right]\,,
\end{eqnarray}
with
\begin{equation}
\alpha^{\prime}_{1}=\alpha_{1}Y_{1}+\alpha_{2}Y_{2}\,,\qquad \alpha^{\prime}_{2}=\alpha_{1}Y_{1}-\alpha_{2}Y_{2}\,.
\end{equation}
Consequently the effect of the complex modulus $\tau$ can be absorbed by couplings in the models without CP-like symmetry. If CP invariance is included in the model, both couplings $\alpha_1$ and $\alpha_2$ would be real, i.e.,
\begin{equation}\label{eq:CP_cons_Mpsip}
\alpha_{1}=\alpha^*_{1}\,, \qquad \alpha_{2}=\alpha^*_{2}\,.
\end{equation}
Then the effect of modular forms can not be removed. The mass matrix and the CP constraint are shown as case $W_{\psi5}$ in the table~\ref{tab:CL_mms}. For the remaining three cases of doublet flavon: $\Phi\sim\bm{2_{2,2}}$,  $\Phi\sim\bm{2_{3,2}}$,  $\Phi\sim\bm{2_{4,2}}$, the predictions for the fermion mass matrices are shown in table~\ref{tab:CL_mms} and are labelled as $W_{\psi6}$, $W_{\psi7}$ and $W_{\psi8}$. The EFG constrains only the singlet modular forms  $Y^{(k_{Y})}_{\bm{1}}(\tau)$ and $Y^{(k_{Y})}_{\bm{1'}}(\tau)$ to appear in the superpotential in these cases, and the modular forms can be absorbed into the coupling constants in the scenario without CP. The representation assignment $\psi\sim\bm{3^1}$, $\psi^c\sim\bm{\bar{3}^1}$ leads to the same mass matrix as that of $\psi\sim\bm{3^0}$, $\psi^c\sim\bm{\bar{3}^0}$. Furthermore, the fermion mass matrix for $\psi\sim\bm{3^0}\, (\bm{3^1})$, $\psi^c\sim\bm{\bar{3}^1}\,(\bm{\bar{3}^0})$ is related to that of $\psi\sim\bm{3^0}$, $\psi^c\sim\bm{\bar{3}^0}$ through the permutation $Y_{1}\leftrightarrow Y_{2}$.

\item[~~(\lowercase\expandafter{\romannumeral3})]  $\psi \sim \bm{3^a}\,(\bm{\bar{3}^a})$, $\psi^c\sim \bm{\bar{3}^b}\,(\bm{3^b})$, $ \Phi\sim \bm{2_{i,m}} $ for $\rho_{\bm{2_{i,m}}}(A)\neq\rho_{\bm{2_{i,m}}}(ST)$

It differs from the case \textbf{ii} above in the representation assignment of the flavon $\Phi\sim \bm{2_{i,m}}$. Although $\Phi$ transforms as a reducible doublet $\bm{2_{i,m}}$ under EFG, the representation matrix of the modular transformation $\rho_{\bm{2_{i,m}}}(ST)$ is not identical with the flavor transformation $\rho_{\bm{2_{i,m}}}(A)$. The corresponding eight combinations of the indices $(i,m)$ are shown in Eq.~\eqref{eq:ST-neq-A}. Similar to previous cases, let us consider a model in which the matter fields $\psi$ and $\psi^c$ are assigned to be EFG triplets $\bm{3^0}$ and $\bm{\bar{3}^0}$ respectively, and the flavon $\Phi=(\phi_{1},\phi_{2})^T$ transforms as $\bm{2_{1,1}}$ under EFG. The $\Delta(27)$ invariant superpotential $\mathcal{W}_{D}$ is of the same form as that of Eq.~\eqref{eq:D27_inv_SP_M2}. For this assignment, the modular transformations of $\Delta(27)$ invariant contractions $\mathcal{O}_{4,5}$ in Eq.~\eqref{eq:def_O4O5} under $S$ and $T$ are
\begin{equation}
\mathcal{O}_4\stackrel{S}{\longrightarrow}\mathcal{O}_5,\qquad \mathcal{O}_4\stackrel{T}{\longrightarrow}\omega \mathcal{O}_5\,,\qquad
\mathcal{O}_5\stackrel{S}{\longrightarrow}\mathcal{O}_4,
\qquad \mathcal{O}_5\stackrel{T}{\longrightarrow}\omega^2\mathcal{O}_4\,.
\end{equation}
One can check that $\mathcal{O}_4$ and $\mathcal{O}_5$ can be arranged into $S_{3}$ doublet $\bm{2}$, i.e.
\begin{equation}
\begin{pmatrix}
\mathcal{O}_4+\omega\mathcal{O}_5\\
-i(\mathcal{O}_4-\omega \mathcal{O}_5)
\end{pmatrix}\sim\bm{2}\,.
\end{equation}
Therefore only the modular form multiplet $Y^{(k_Y)}_{\bm{2}}=\left(Y_3, Y_4\right)^T$ is relevant, and the eclectic invariant superpotential is given by
\begin{eqnarray}
\mathcal{W}_{D}=\frac{\alpha}{\Lambda}\left[Y_{3}\left(\mathcal{O}_4+\omega\mathcal{O}_5\right) -iY_4\left(\mathcal{O}_4-\omega\mathcal{O}_5\right)\right]H_{u/d}\,.
\end{eqnarray}
Then one can read out the mass matrix of fermion $\psi$ as follow,
\begin{equation}\label{eq:Ch_mass3}
M^{\prime\prime}_{\psi}= \frac{\alpha v_{u,d}}{\Lambda}\left(
\begin{array}{ccc}
0 & \omega\phi_{2} (Y_{3}+i Y_{4}) & \phi_{1}  (Y_{3}-i Y_{4}) \\
\phi_{1}  (Y_{3}-i Y_{4}) & 0 & \omega\phi_{2} (Y_{3}+i Y_{4}) \\
\omega\phi_{2} (Y_{3}+i Y_{4}) & \phi_{1} (Y_{3}-i Y_{4}) & 0
\end{array}\right)\,.
\end{equation}
There is only an overall parameter $\alpha$ which can be set to real whenever CP-like symmetry is included or not. The superpotential and fermion mass matrices can be determined in a similar way for the other seven possible assignments of the doublet flavon $\Phi$, and the results are summarized in table~\ref{tab:CL_mms}. It is remarkable that only the doublet modular forms contributes to the Yukawa couplings. If the matter fields $\psi$ and $\psi^c$ transform as $\bm{3^1}$ and $\bm{\bar{3}^1}$ respectively under EFG while the representation assignment of the flavon $\Phi$ remains unchanged, one would obtain the same superpotential and mass matrix. If one assigns the matter fields $\psi\sim\bm{3^a}$ and $\psi^{c}\sim\bm{\bar{3}^b}$ with $a+b=1$, the $\Delta(27)$ invariant combinations $\mathcal{O}_{4,5}$ transform under $S$ and $T$ as follow,
\begin{equation}
\mathcal{O}_4\stackrel{S}{\longrightarrow}-\mathcal{O}_5,\qquad \mathcal{O}_4\stackrel{T}{\longrightarrow}-\omega \mathcal{O}_5\,,\qquad \qquad
\mathcal{O}_5\stackrel{S}{\longrightarrow}-\mathcal{O}_4,
\qquad \mathcal{O}_5\stackrel{T}{\longrightarrow}-\omega^2\mathcal{O}_4\,,
\end{equation}
for $\Phi=(\phi_{1},\phi_{2})^T\sim \bm{2_{1,1}}$. Consequently we can organize $\mathcal{O}_4$ and $\mathcal{O}_5$ into a doublet of the modular symmetry $S_3$,
\begin{equation}
\begin{pmatrix}
\mathcal{O}_4-\omega\mathcal{O}_5\\
-i(\mathcal{O}_4+\omega \mathcal{O}_5)
\end{pmatrix}\sim\bm{2}\,.
\end{equation}
The corresponding fermion mass matrix can be obtained from that of Eq.~\eqref{eq:Ch_mass3} through the replacement $\phi_{2}\rightarrow -\phi_{2}$. Moreover, the fermion mass matrix would be transposed if the representation assignments of $\psi$ and $\psi^{c}$ are interchanged, i.e. $\psi\sim\bm{\bar{3}^b}$ and $\psi^{c}\sim\bm{3^a}$.

\end{description}

In above, we have performed a comprehensive analysis for Dirac mass terms which are invariant under the EFG $\Delta(27)\rtimes S_{3}$. If a field $\psi^c$ is a SM singlet, for instance $\psi^c$ can be the right-handed neutrinos or the combination of left-handed leptons and Higgs, the Majorana mass term is allowed and it is of the following form
\begin{equation}
\mathcal{W}_{M}=\sum_{\bm{r},s}c_{\bm{r},s}\left(Y^{(k_Y)}_{\bm{r}}\Phi \psi^{c}\psi^c\right)_{\bm{1^{0}_{0,0}},s}\,.
\end{equation}
By dropping the antisymmetric contributions, the Majorana mass matrix of $\psi^c$ can be easily obtained from the cases in which $\psi$ and $\psi^c$ transform in the same way under EFG. For the triplet assignment $\psi^c\sim \bm{3^a}$ or $ \bm{\bar{3}^a}$, the Majorana mass matrix of $\psi^c$ can be obtained from mass matrices of $\mathcal{W}_{\psi1}\sim \mathcal{W}_{\psi4}$ in table~\ref{tab:CL_mms} by taking $\alpha_{3}=0$ and $v_{u,d}/\Lambda=1$.

\section{\label{sec:example-model-EFG}An example model based on the EFG $\Delta(27)\rtimes S_{3}$}

In the following, we shall present a lepton model with the EFG $\Delta(27)\rtimes S_{3}$, and a $Z_3$ symmetry is employed to forbid the unwanted operators. We formulate our model in the framework of type-I seesaw mechanism with three RH neutrinos. We assign the three generations of LH lepton doublets $L$ and of RH charged leptons to two triplet  $\bm{\bar{3}}$ of $\Delta(27)$, while the RH neutrinos $N^c$ furnish a three-dimensional irreducible representation $\bm{3}$ of $\Delta(27)$. The Higgs doublets $H_u$ and $H_d$ are assumed to transform trivially under $\Delta(27)$ and their modular weights are vanishing. The fields of the model and their classification under the EFG  $\Delta(27)\rtimes S_{3}$ and $Z_3$ are summarized in table~\ref{tab:model-fields}. In the traditional discrete flavor symmetry approach, a number of scalar fields (flavons) are generally required to break the flavor symmetry~\cite{King:2017guk,Feruglio:2019ybq,Altarelli:2005yx}. The flavons are standard model singlets, and yet they transform non-trivially under the flavor symmetry group. The VEVs of flavons should be aligned along particular directions in the flavor space, and they typically break the traditional flavor symmetry down to certain Abelian subgroups. In the present model, we introduce three flavons $\phi$, $\varphi$ and $\chi$ which transform as $\bm{\bar{3}}$, $\bm{2_{2}}$ and $\bm{3}$ under $\Delta(27)$, respectively. From table~\ref{tab:model-fields}, we find that only the irreducible representations $\bm{1^{0}_{0,0}}$, $\bm{2_{2,0}}$, $\bm{3^{0}}$ and $\bm{\bar{3}^{0}}$ are introduced in our model and they are mapped to their complex conjugates under the action of the automorphism $u_{K^{*}}$, as shown in Eq.~\eqref{eq:uKstar-rep}. Hence the CP-like transformation corresponding to $u_{K^{*}}$  can be consistently imposed as a symmetry which leads to physical CP conservation in our model. In this model, we assume that the VEV of flavon $\phi$ breaks the traditional flavor symmetry $\Delta(27)$ down to $Z^{BAB^2}_{3}\equiv\left\{1, BAB^2, BA^2B^2\right\}$ while the subgroup $Z^{TST}_{2}\equiv\left\{1, TST\right\}$ of $S_{3}$ is preserved  by vacuum of flavons $\varphi$ and $\chi$. Hence the flavons $\phi$, $\varphi$ and $\chi$ will develop VEVs along the following directions:
\begin{equation}\label{eq:seesaw_VEVs}
\langle\phi\rangle=(1,\omega,\omega^2)^Tv_{\phi}, \qquad	\langle\varphi\rangle=(1,1)^Tv_{\varphi}, \qquad \langle\chi\rangle=(1,x,1)^Tv_{\chi}\,,
\end{equation}
where the parameter $x$ is undetermined and generally complex. In the following numerical analysis, we shall take $x$ to be real for simplicity. In our model, the phases of the VEVs $v_{\phi}$, $v_{\varphi}$ and $v_{\chi}$ are unphysical since they are the overall phases of the charged lepton mass matrix, neutrino Dirac mass matrix and neutrino Majorana mass matrix, respectively. It is notoriously difficult to realize the vacuum alignment of flavon, one has to construct rather complicated flavon potential, and additional symmetry such as $U(1)_R$ and new fields are generally required. Both flavons and complex modulus are present in the paradigm of EFG, and the interplay between them makes the dynamical determination of the flavon vacuum alignment even more difficult. Moreover, the modular invariant potential of the modulus $\tau$ contains a lot of independent terms~\cite{Novichkov:2022wvg,Cvetic:1991qm,Gonzalo:2018guu}. Hence it is very challenging to determine the modulus VEV from a dynamical principle.
We will not address the vacuum alignment problem here, and we rely on some unknown vacuum selection mechanism.

\begin{table}[t!]
\renewcommand{\tabcolsep}{2.05mm}
\renewcommand{\arraystretch}{1.1}
\centering
\begin{tabular}{|c|c|c|c|c|c||c|c|c|}\hline \hline
Fields &   $L$    &  $E^{c}$  & $N^c$   &  $H_{u}$ & $H_{d}$ &  $\phi$  &  $\varphi$  &  $\chi$   \\ \hline
		
$\text{SU}(2)_L\times \text{U}(1)_{Y}$ & $(\bm{2},-\frac{1}{2})$ & $(\bm{1},1)$ & $(\bm{1},0)$ & $(\bm{2},\frac{1}{2})$ & $(\bm{2},-\frac{1}{2})$ &  $(\bm{1},0)$ &  $(\bm{1},0)$ &  $(\bm{1},0)$ \\ \hline

$\Delta(27)\rtimes S_{3}$ & $\bm{\bar{3}^{0}}$ & $\bm{\bar{3}^{0}}$ & $\bm{3^{0}}$ & $\bm{1^{0}_{0,0}}$  & $\bm{1^{0}_{0,0}}$ & $\bm{\bar{3}^{0}}$ & $\bm{2_{2,0}}$ & $\bm{3^{0}}$   \\ \hline
		
Modular weight & $0$ & $0$ &$0$ &  $0$ & $0$ & $6$ & $4$ & $0$   \\  \hline
		
$Z_3$ & $\omega$ & $\omega$ & $1$ & $1$ & $1$ & $\omega$ &  $\omega^2$  & $1$  \\ \hline \hline
		
\end{tabular}
\caption{\label{tab:model-fields}Transformation properties of matter fields, Higgs doublets and flavons under the EFG $\Delta(27)\rtimes S_{3}$ and the auxiliary symmetry $Z_{3}$.	}
\end{table}

\begin{figure}[t!]
\centering
\begin{tabular}{c}
\includegraphics[width=0.95\linewidth]{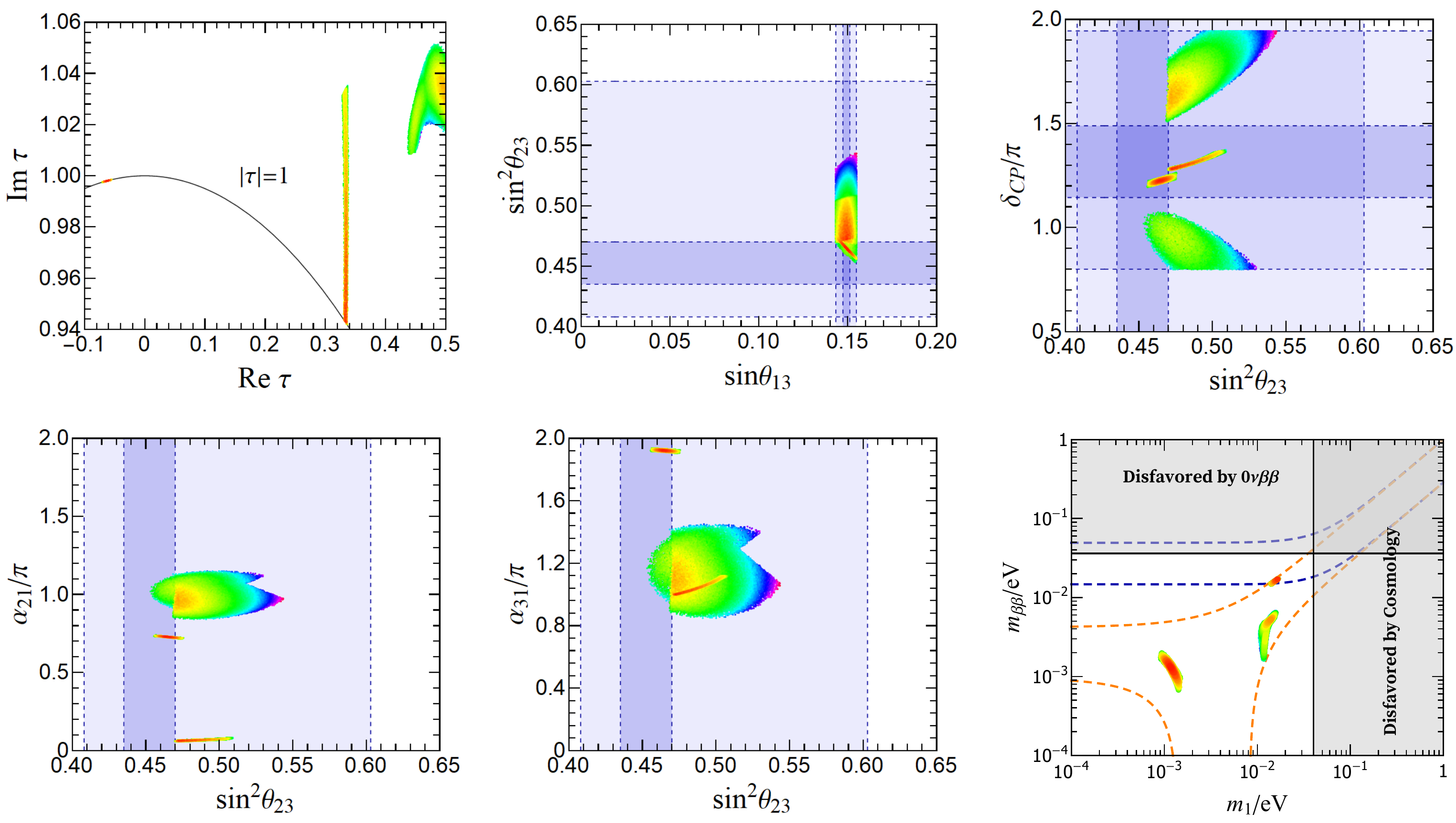}\\
\includegraphics[width=0.38\linewidth]{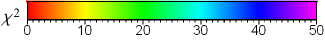}\\
\end{tabular}
\caption{\label{fig:M_seesaw} The predicted correlations among the input free parameters, neutrino mixing angles, CP violation phases and the effective mass in neutrinoless double beta decay.}
\end{figure}

Given the field content and the symmetry assignment in table~\ref{tab:model-fields}, we find that the superpotential for the lepton masses, which is invariant under the EFG $\Delta(27)\rtimes S_{3}$, is of the form
\begin{equation}
\mathcal{W}=\mathcal{W}_{l}+\mathcal{W}_{\nu}\,,
\end{equation}
with
\begin{eqnarray}
\nonumber \mathcal{W}_{l}&=&\frac{\alpha}{\Lambda}  \left(E^c L\phi Y^{(6)}_{\bm{1}}\right)_{(\bm{1_{0,0}},\bm{1}),1} H_d+\frac{\beta}{\Lambda}  \left(E^c L\phi Y^{(6)}_{\bm{1}}\right)_{(\bm{1_{0,0}},\bm{1}),2} H_d+\frac{\gamma}{\Lambda}  \left(E^c L\phi Y^{(6)}_{\bm{1^\prime}}\right)_{(\bm{1_{0,0}},\bm{1})} H_d\,, \\
\label{eq:seesaw_SP}\mathcal{W}_{\nu}&=&\frac{h}{\Lambda}  \left(N^c L\varphi Y^{(2)}_{\bm{2}}\right)_{(\bm{1_{0,0}},\bm{1})} H_u+\frac{g_{1}}{2}\left(N^cN^c\chi \right)_{(\bm{1_{0,0}},\bm{1}),1}+\frac{g_{2}}{2}\left(N^cN^c\chi \right)_{(\bm{1_{0,0}},\bm{1}),2}\,,
\end{eqnarray}
where $\mathcal{W}_{l}$ is the Yukawa superpotential of the charged leptons, $\mathcal{W}_{\nu}$ is the neutrino superpotential in type-I seesaw mechanism. We impose the CP-like symmetry in the model, thus the couplings should fulfill the following relations:
\begin{equation}
\beta=\frac{\omega^2}{2}(\sqrt{3}\alpha^*-\alpha)\,, \qquad  \text{Arg}(\gamma)=\frac{\pi}{12}\,\,\,\,(\text{mod} \,\pi)\,, \qquad
g_{2}=\frac{\omega}{2}(\sqrt{3}g_{1}^*-g_{1})\,.
\end{equation}
From the results in table~\ref{tab:CL_mms}, we find that the charged lepton mass terms correspond to the case of $\mathcal{W}_{\psi3}$. With the vacuum configuration of $\phi$ in Eq.~\eqref{eq:seesaw_VEVs}, the charged lepton mass matrix is given by
\begin{eqnarray}
\nonumber &&M_{l}=\frac{ v_dv_{\phi}}{\Lambda}\left[Y^{(6)}_{\bm{1}}\left(
	\begin{array}{ccc}
		\alpha  & \beta  \omega  & \beta  \omega ^2 \\
		\beta  \omega  & \alpha  \omega ^2 & \beta  \\
		\beta  \omega ^2 & \beta  & \alpha  \omega  \\
	\end{array}
	\right)+\gamma Y^{(6)}_{\bm{1^\prime}}\left(
	\begin{array}{ccc}
		0 & -  \omega  &   \omega ^2 \\
		\omega  & 0 & -1 \\
		-  \omega ^2 & 1  & 0 \\
	\end{array}
	\right)\right]\,.
\end{eqnarray}
We see that the charged lepton sector involves a single flavon $\phi$ whose VEV preserves the subgroup $Z_3^{BAB^2}$. Therefore the hermitian combination $M^\dagger_{l}M_{l}$ is invariant under the transformation $L\rightarrow\rho_{\bm{\bar{3}}}(BAB^2)L$ and it satisfies the identity $\rho^\dagger_{\bm{\bar{3}}}(BAB^2)M^\dagger_{l}M_{l}\rho_{\bm{\bar{3}}}(BAB^2)=M^\dagger_{l}M_{l}$. As a consequence, $M^\dagger_{l}M_{l}$ is diagonalized by the following constant unitary matrix
\begin{equation}\label{eq:Ue1}
U_{l}=\frac{1}{\sqrt{3}}\left(
\begin{array}{ccc}
1 ~& -\omega^2 ~& 1 \\
\omega^2 ~& -1 ~& 1 \\
\omega  ~& -\omega  ~& 1
\end{array}
\right)\,,
\end{equation}
with $U_{l}^{\dagger} M^\dagger_{l}M_{l}U_{l}=\text{diag}(m^2_e,m^2_{\mu},m^2_{\tau})$ and the charged lepton masses
\begin{eqnarray}
\nonumber && m_{e}=\left|(\alpha +2\beta)Y^{(6)}_{\bm{1}}\right|\frac{ v_{\phi}v_d}{\Lambda} , \\
\nonumber && m_{\mu}=\left|(\alpha- \beta) Y^{(6)}_{\bm{1}}+\sqrt{3}i\gamma Y^{(6)}_{\bm{1^\prime}} \right|\frac{ v_{\phi}v_d}{\Lambda}, \\
&& m_{\tau}=\left|(\alpha- \beta)Y^{(6)}_{\bm{1}}-\sqrt{3}i\gamma Y^{(6)}_{\bm{1^\prime}} \right|\frac{ v_{\phi}v_d}{\Lambda}\,.
\end{eqnarray}
In the neutrino sector, we see that the neutrino Dirac mass term corresponds to $\mathcal{W}_{\psi11}$ of table~\ref{tab:CL_mms} and the neutrino Majorana mass terms can be obtained from $\mathcal{W}_{\psi1}$ by taking $\alpha_{3}=0$. The Dirac neutrino mass matrix and the RH Majorana neutrino  mass matrix read as
\begin{eqnarray}
\nonumber &&M_{D}=\frac{hv_{\varphi}v_u}{\Lambda}\left(
\begin{array}{ccc}
0 ~& Y^{(2)}_{\bm{2},1}-i Y^{(2)}_{\bm{2},2} ~& \omega (Y^{(2)}_{\bm{2},1}+i Y^{(2)}_{\bm{2},2}) \\
 Y^{(2)}_{\bm{2},1}+i Y^{(2)}_{\bm{2},2} ~& 0 ~& \omega (Y^{(2)}_{\bm{2},1}-i Y^{(2)}_{\bm{2},2}) \\
\omega^2 (Y^{(2)}_{\bm{2},1}-i Y^{(2)}_{\bm{2},2}) ~& \omega^2(Y^{(2)}_{\bm{2},1}+i Y^{(2)}_{\bm{2},2})  ~& 0 \\
\end{array}
\right)\,, \\
\label{eq:M1_mass} && M_{N}=v_{\chi}\left(
\begin{array}{ccc}
g_{1} ~& g_{2} ~& g_{2} x \\
g_{2} ~& g_{1} x ~& g_{2} \\
g_{2} x ~& g_{2} ~& g_{1} \\
\end{array}
\right)\,,
\end{eqnarray}
where $Y^{(k_Y)}_{\bm{r},i}$ denotes the $i$th component of the modular form multiple $Y^{(k_Y)}_{\bm{r}}(\tau)$. It is notable that $M_D$ is completely determined by the modulus $\tau$ up to the overall scale $hv_u v_{\varphi}/\Lambda$. The light neutrino mass matrix is then given by the seesaw formula $M_{\nu}=-M^T_{D}M^{-1}_{N}M_{D}$. As explained early, the complex modulus $\tau$ is treated as spurion and its value is freely varied in the fundamental domain $\mathcal{D}=\left\{\tau\in\mathbb{C}\big|-1/2\leq\Re(\tau)\leq1/2, |\tau|\geq1\right\}$ to adjust the agreement with the experimental data. Taking into account CP-like symmetry, we see that the lepton mass matrices depend on the following six dimensionless real parameters $\Re{\tau}$, $\Im{\tau}$, $\text{arg}{\left(\alpha\right)}$, $\gamma/|\alpha|$, $\text{arg}{\left(g_{1}\right)}$, $x$ and two overall scales $\big|\alpha v_{d}v_{\phi}/\Lambda\big|$ and $\big|h^2v^2_uv^2_{\varphi}/(g_{1}v_{\chi}\Lambda^2)\big|$ which can be determined by the measured electron mass and the solar neutrino mass square difference $\Delta m^2_{21}$. In order to quantitatively determine how well the model can describe the experimental data on lepton masses and mixing parameters, we perform a conventional $\chi^2$ analysis and the $\chi^2$ function is constructed with the data listed in table~\ref{tab:bf-1sigma-3sigma-data}. We search for the minimum of the $\chi^2$ function to obtain the best fit values of the free input parameters as well as the predictions for the lepton masses and mixing parameters.  It is remarkable that the model can accommodate both NO and IO neutrino mass spectrum. A good agreement between the model and the experimental data can be achieved for the following values of free parameters:
\begin{table}[t!]
\centering
\begin{tabular}{|c|c|c||c|c|}\hline \hline
\multirow{2}{*}{Observables}  &  	\multicolumn{2}{c|}{NO}   &      \multicolumn{2}{c|}{IO}       \\ \cline{2-3} \cline{4-5}
	
& $\text{bf}\pm1\sigma$  & $3\sigma$ region & $\text{bf}\pm1\sigma$ & $3\sigma$ region   \\ \hline

&   &  &  &    \\[-0.150in]
	
$\sin^2\theta_{13}$ & $0.02225^{+0.00056}_{-0.00059}$ & $[0.02052,0.02398]$ & $0.02223^{+0.00058}_{-0.00058}$ &  $[0.02048,0.02416]$  \\ [0.050in]
	
$\sin^2\theta_{12}$ & $0.303^{+0.012}_{-0.012}$ & $[0.270,0.341]$ & $0.303^{+0.012}_{-0.011}$ & $[0.270,0.341]$  \\ [0.050in]
	
$\sin^2\theta_{23}$  & $0.451^{+0.019}_{-0.016}$ & $[0.408,0.603]$ & $0.569^{+0.016}_{-0.021}$  & $[0.412,0.613]$   \\ [0.050in]
	
$\delta_{CP}/\pi$  & $1.289^{+0.20}_{-0.14}$ &  $[0.8,1.944]$  & $1.533^{+0.122}_{-0.161}$ & $[1.08,1.911]$  \\ [0.050in]

$\Delta m^2_{21}/\Delta m^2_{3\ell}$  &  $0.0294^{+0.00088}_{-0.00088}$   & $[0.0263,0.0331]$ & $-0.0296^{+0.00064}_{-0.00088}$ & $[-0.0331,-0.0263]$  \\ [0.050in]

$m_e/m_{\mu}$  & $0.004737^{+0.00004}_{-0.00004}$   & --- & $0.004737^{+0.00004}_{-0.00004}$ & --- \\ [0.050in]

$m_{\mu}/m_{\tau}$  &  $0.05876^{+0.000465}_{-0.000465}$  & --- & $0.05876^{+0.000465}_{-0.000465}$ & --- \\ [0.050in] \hline \hline
	
\end{tabular}
\caption{\label{tab:bf-1sigma-3sigma-data}
The global best fit values and $1\sigma$ ranges and $3\sigma$ ranges of the lepton mixing parameters and mass ratios, where the experimental data and errors of the lepton mixing parameters and neutrino masses are obtained from NuFIT5.2 with SK atmospheric data~\cite{Esteban:2020cvm}, the charged lepton mass ratios are taken from~\cite{Antusch:2013jca} with $M_{\text{SUSY}}=10\,\text{TeV}$ and $\tan\beta=5$. Note that $\Delta m^2_{3\ell}=\Delta m^2_{31}>0$ for normal ordering (NO) and $\Delta m^2_{3\ell}=\Delta m^2_{32}<0$ for inverted ordering (IO). }
\end{table}
\begin{eqnarray}
\nonumber &&	\Re\langle\tau\rangle=-0.0621\, (0.496), \qquad \Im\langle\tau\rangle=0.998\,(0.965), \qquad \text{arg}{\left(\alpha\right)}=0.100\pi \,(0.0833\pi), \\
\nonumber && \gamma/|\alpha|=0.207\, (-0.778), \qquad \text{arg}{\left(g_{1}\right)}=0.119 \,(0.0111), \qquad x=1.882(0.987), \\
&& 	|\alpha| v_{d}v_{\phi}/\Lambda=739.4\,\text{GeV}\, (159.3\,\text{GeV})\,, \qquad	h^2v^2_uv^2_{\varphi}/(|g_{1}|v_{\chi}\Lambda^2)=0.815\,\text{eV} \,(1.803\,\text{eV})\,,
\end{eqnarray}
for NO (IO). The minimum value of the $\chi^2$ function is found to be $\chi^2_{\text{min}}=0.751 (4.132)$. The best fit values of the lepton masses and mixing parameters are given by
\begin{eqnarray}
\nonumber &&\hskip-0.25in \sin^2\theta_{13}=0.02233\, (0.02216), \qquad \sin^2\theta_{12}=0.3041 \,(0.3025), \qquad \sin^2\theta_{23}=0.4645\, (0.5945), \\
\nonumber &&\hskip-0.25in  \delta_{CP}=1.226\pi\, (1.498\pi)\,,
\qquad \alpha_{21}=0.729\pi \,(1.998\pi),\qquad \alpha_{31}=1.923\pi\,(0.994\pi),\\
\nonumber &&\hskip-0.25in m_1=1.173\,\text{meV} \,(72.26\,\text{meV}) ,\quad m_2=8.676\,\text{meV}\, (72.77\,\text{meV})\,, \quad   m_3=50.13\,\text{meV} \,(52.84\,\text{meV}), \\
\nonumber &&\hskip-0.25in \sum^3_{i=1}m_{i}=59.98\,\text{meV} \,\left(197.9\,\text{meV}\right), \qquad 	m_{\beta\beta}=1.298\,\text{meV}\, (71.98\,\text{meV})\,,\\
&&\hskip-0.25in m_{e}=0.511\,\text{MeV}\, (0.511\,\text{MeV}) , ~~  m_{\mu}=107.9\,\text{MeV} \,(107.9\,\text{MeV}) , ~~  m_{\tau}=1.837\,\text{GeV} \,(1.836\,\text{GeV}) \,.~~~~~~~
\end{eqnarray}
We see that the value of $\tau$ is very close to the self-dual point $\tau=i$ for NO. Notice that the sum of the three neutrino mass is determined to be  $59.98\,\text{meV}$ for NO, and it is much below the most stringent bound $\sum m_{\nu}<120\,\text{meV}$ from Planck~\cite{Planck:2018vyg}. In the case of IO, we find $\sum^3_{i=1}m_{i}=197.9\,\text{meV}$ is still allowed by the conservative upper bound $\sum\,m_{\nu}<600\,\text{meV}$ although it is above the stringent bound $\sum\,m_{\nu}<120\,\text{meV}$~\cite{Planck:2018vyg}. Moreover, the best fit value of the effective Majorana mass is $m_{\beta\beta}=1.298\,\text{meV}\,(71.98\,\text{meV})$ which is compatible with the latest result $m_{\beta\beta}<36-156$ meV of KamLAND-Zen~\cite{KamLAND-Zen:2022tow}. The prediction of neutrinoless double decay for IO can be tested by the next generation experiments such as LEGEND~\cite{LEGEND:2017cdu} and nEXO~\cite{nEXO:2017nam} which are expected to explore the full IO region. In order to show the viability and predictions of the model, we shall numerically scan over the parameter space of the model and require all the observables lie in their experimentally preferred $3\sigma$ regions. Then some interesting correlations among the input parameters and observables are obtained and the corresponding correlations among different observables are shown in figure~\ref{fig:M_seesaw} for the NO spectrum.

As can be seen from table~\ref{tab:model-fields}, all the lepton fields are assigned to triplets of $\Delta(27)$, and the model does not contain a doublet flavon invariant under the extra symmetry $Z_3$. From the general results about the K\"ahler potential in section~\ref{subsubsec:kahler1},  we know that there is no NLO corrections to the minimal K\"ahler potential in this model. The off-diagonal contributions of the K\"ahler metric arise at NNLO and they are suppressed by $|\langle\Phi\rangle|^2/\Lambda^2$, where $\Phi$ denotes any flavon of the model. Hence the contributions of the K\"ahler potential to the lepton masses and mixing parameters are suppressed by $|\langle\Phi\rangle|^2/\Lambda^2$, they are small enough to be negligible.

\section{\label{sec:conclusion} Conclusion and outlook}

Usually the minimal K\"ahler potential is adopted in modular symmetry model building. However, it is not the most general one compatible with modular symmetry, the non-minimal and flavor-dependent terms are allowed. How to restrict the K\"ahler is an open question in modular symmetry. The top-down approach to modular flavor symmetry in string inspired constructions can give rise to both traditional flavor symmetry and modular symmetry. This results in the idea of EFG which is the nontrivial product the modular and traditional flavor symmetries. The EFG can severely restrict both K\"ahler potential and superpotential.

In the present work, we have studied the traditional flavor group $\Delta(27)$ extended by the finite modular group $\Gamma_2\cong S_{3}$, and the resulting EFG is $\Delta(27)\rtimes S_{3}$. Note that $S_3$ is a subgroup of the automorphism group of $\Delta(27)$. The modular transformations $S$, $T$, $TST$ correspond to outer automorphisms of $\Delta(27)$ while others are inner automorphisms. In order to consistently combine the modular symmetry $S_3$ with the traditional flavor symmetry $\Delta(27)$, we find that the eight nontrivial singlet representations of $\Delta(27)$ should be arranged into four reducible doublets $\bm{2_{i}}$ ($i=1,\cdots,4$) and the remain three irreducible representations $\bm{1_{0,0}}$, $\bm{3}$ and $\bm{\bar{3}}$ need not be extended. Considering the modular symmetry, we find that the EFG  $\Delta(27)\rtimes S_{3}$ has two one-dimensional representations $\bm{1^a_{0,0}}$, twelve two-dimensional representations $\bm{2_{i,m}}$, four three-dimensional representations $\bm{3^a}$ and $\bm{\bar{3}^a}$, and two six-dimensional representations $\bm{6}$ and $\bm{\bar{6}}$. Furthermore, we also extend the EFG $\Delta(27)\rtimes S_{3}$ to include the CP-like symmetry and give the explicit form of the CP transformation matrices.

We have performed a comprehensive analysis of the superpotential and K\"ahler potential which are invariant under the action of the EFG $\Delta(27)\rtimes S_{3}$. We find that the K\"ahler potential is under control when the chiral superfields of quarks/leptons are assigned to trivial singlet or triplet of $\Delta(27)$, and the minimal K\"ahler potential is reproduced at leading order. The flavor-dependent terms of K\"ahler potential are suppressed by $\langle\Phi\rangle^2/\Lambda^2$ unless the model contain a $\Delta(27)$ doublet flavon invariant under auxiliary cyclic symmetries, where $\Phi$ denotes a generic flavon. On the other hand, the K\"ahler metric is diagonal but flavor-dependent and the contributions to flavor observables are not negligible, if the quark/lepton fields are assigned to a singlet plus a doublet of $\Delta(27)$. Moreover, we analyze the superpotential of fermion mass for various possible representation assignments of matter fields and flavons, the predictions for the fermion mass matrix are summarized in table~\ref{tab:CL_mms}.

Furthermore, we propose a bottom-up model for lepton masses and mixing based on the EFG $\Delta(27)\rtimes S_{3}$. In contrast with the top-down models, we freely assign the representations and modular weights of the fields, although the modular transformation is fixed by the transformation under $\Delta(27)$ flavor symmetry. We introduce an extra symmetry $Z_{3}$ to forbid the undesired operator, two triplet flavons and one doublet flavon are introduced to break the EFG. It is assumed that the subgroup $Z_3^{ BAB^2}$ of $\Delta(27)$ and $Z^{TST}_{2}$ of $S_3$ are preserved by the VEVs of flavons in the charged lepton and neutrino sectors respectively. When CP-like symmetry consistent with the EFG $\Delta(27)\rtimes S_{3}$ is imposed, the six lepton masses and six mixing parameters depend on eight real input parameters. A comprehensive numerical analysis are performed, we find that the model is in excellent agreement with experimental data for certain values of free parameters. The predictions for the three neutrino mass sum and the effective mass $m_{\beta\beta}$ of neutrinoless double beta decay are safely below the present upper limit.

In this work, we explicitly show it is not obligatory that all elements of finite modular group need correspond to outer automorphisms of the traditional flavor symmetry group. Even some modular symmetry elements are inner automorphisms of the flavor symmetry group, we could still get nontrivial results and non-singlet modular forms could appear in the Yukawa couplings if the corresponding modular transformation matrices do not coincide with the flavor symmetry transformations.

\section*{Acknowledgements}

CCL is supported by the National Natural Science Foundation of China under Grant Nos. 12005167, 12247103, Natural Science Basic Research Program of Shaanxi (Program No. 2024JC-YBQN-0004) and the Young Talent Fund of Association for Science and Technology in Shaanxi, China. GJD is supported by the National Natural Science Foundation of China under Grant Nos.~11975224, 11835013.

\newpage

\section*{Appendix}

\setcounter{equation}{0}
\renewcommand{\theequation}{\thesection.\arabic{equation}}

\begin{appendix}

\section{\label{sec:Delta27_group}Traditional flavor symmetry $\Delta(27)$}

The group structure of the traditional flavor group $\Delta(27)$ is $\Delta(27)\cong(Z_3\times Z_3)\rtimes Z_3$ which is a
non-Abelian group of order 27 with GAP ID $[27,3]$ in GAP~\cite{GAP,SmallGroups}. In detail, $\Delta(27)$ can be generated by two generators
$A$ and $B$ obeying the relations
\begin{equation}\label{eq:D27_Mul_ruls}
A^3 = B^3 =  (AB)^3 = (AB^2)^3 =1\,.
\end{equation}
The center of the traditional flavor group $\Delta(27)$, denoted by $Z(\Delta(27))$, is of the following form
\begin{equation}
Z(\Delta(27))=\{1, BAB^{2}A^2,  ABA^{2}B^2\}\,,
\end{equation}
which is a normal Abelian $Z_{3}$ subgroup of $\Delta(27)$. The 27 group elements of $\Delta(27)$ can be divided into the eleven conjugacy classes as follows
\begin{eqnarray}
\nonumber &1C_1=\{1\}\,, \qquad &3C^{(1)}_3=\left\{A^2B^2,B^2A^2,AB^2A\right\}\,,\\
\nonumber 	&3C^{(2)}_3=\left\{A^2B,ABA,BA^2\right\}\,, \qquad & 3C^{(3)}_3=\left\{A,BAB^2,B^2AB\right\}\,, \\
\nonumber & 3C^{(4)}_3=\left\{AB^2,BAB,B^2A\right\}\,, \qquad & 3C^{(5)}_3=\left\{AB,BA,A^2BA^2\right\}\,, \\
\nonumber & 3C^{(6)}_3=\left\{A^2,B^2A^2B,BA^2B^2\right\}\,, \qquad &3C^{(7)}_3=\left\{B^2,AB^2A^2,A^2B^2A\right\}\,, \\
\label{eq:D27CC}&3C^{(8)}_3=\left\{B,A^2BA,ABA^2\right\}\,,\qquad & 1C^{(1)}_3=\left\{BAB^2A^2\right\}\,, \quad 1C^{(2)}_3=\left\{ABA^2B^2\right\}\,,
\end{eqnarray}
where $kC_{n}$ denotes a conjugacy class which contains $k$ elements with order $n$. Since the number of conjugacy class is equal to the number of irreducible representation, $\Delta(27)$ has eleven inequivalent irreducible representations which contain nine singlets labeled as $\bm{1_{(r,s)}}$ ($r,s=0,1,2$) and two triplets labeled as $\bm{3}$ and $\bm{\bar{3}}$. In our working basis, the
explicit forms of the generators $A$ and $B$ in the eleven irreducible representations of $\Delta(27)$ are as follows
\begin{eqnarray}
\nonumber \bm{1_{r,s}}&:&~~\rho_{\bm{1_{r,s}}}(A)=\omega^{r}, \qquad \rho_{\bm{1_{r,s}}}(B)=\omega^{s} \,, \quad \text{with} \quad r,s=0,1,2\,, \\
\nonumber  \bm{3}&:&~~\rho_{{\bf 3}}(A)=\left(\begin{array}{ccc}
		0&1&0\\
		0&0&1\\
		1&0&0
\end{array}\right),  \qquad
\rho_{{\bm{3}}}(B)=\left(\begin{array}{ccc}
		1 & 0 & 0 \\
		0 & \omega & 0 \\
		0 & 0 & \omega^{2}
\end{array}\right) \,, \\
\label{eq:Delta27_irre}	 \bm{\bar{3}}&:&~~\rho_{\bm{\bar{3}}}(A)=\left(\begin{array}{ccc}
		0&1&0\\
		0&0&1\\
		1&0&0
	\end{array}\right),  \qquad
\rho_{\bm{\bar{3}}}(B)=\left(\begin{array}{ccc}
		1 & 0 & 0 \\
		0 & \omega^{2} & 0 \\
		0 & 0 & \omega
\end{array}\right) \,.
\end{eqnarray}
We see that $\bm{3}$ and $\bm{\bar{3}}$ are complex conjugate to each other. For a triplet $\Phi=(\phi_1, \phi_2, \phi_3)^T\sim\bm{3}$, it complex conjugate $\Phi^{*}=(\phi^{*}_1, \phi^{*}_2, \phi^{*}_3)^T$ transforms as $\bm{\bar{3}}$ under $\Delta(27)$. The character table of $\Delta(27)$ and the transformation properties of conjugacy classes and irreducible representations under the actions of outer automorphisms $u_{S}$, $u_{T}$ and $u_{K_*}$ are summarized in table~\ref{tab:character_Delta27}.  In table~\ref{tab:character_Delta27}, the second line indicates representatives of the eleven conjugacy classes. From the character table, the Kronecker products between different irreducible representations read as
\begin{eqnarray}
\nonumber &&\bm{1_{r,s}}\otimes \bm{1_{r^\prime,s^\prime}}=\bm{1_{[r+r^\prime],\,\,[s+s^\prime]}} , \qquad  \bm{1_{r,s}}\otimes \bm{3}=\bm{3}, \qquad \bm{1_{r,s}}\otimes \bm{\bar{3}}=\bm{\bar{3}}\,,  \\
\label{eq:KP_Delta27}	&& \bm{3}\otimes \bm{3}=\bm{\bar{3}_{S,1}}\oplus\bm{\bar{3}_{S,2}}\oplus\bm{\bar{3}_{A}} \,,   \qquad  \bm{\bar{3}}\otimes\bm{\bar{3}}=\bm{3_{S,1}}\oplus\bm{3_{S,2}}\oplus\bm{3_{A}} \,,  \qquad \bm{3}\otimes \bm{\bar{3}}=\sum^{2}_{r,s=0} \bm{1_{r,s}}\,,
\end{eqnarray}
where $r,s,r^\prime,s^\prime=0,1,2$ and integer $[n]$  stands for $n$ mod 3.

\begin{table}
\centering
\begin{tabular}{c|ccccccccccc}	\\
& \tabnode{$1C_{1}$} & \tabnode{$3C_{3}^{(1)}$} & \tabnode{$3C_{3}^{(2)}$} & \tabnode{$3C_{3}^{(3)}$} & \tabnode{$3C_{3}^{(4)}$} & \tabnode{$3C_{3}^{(5)}$} & \tabnode{$3C_{3}^{(6)}$} & \tabnode{$3C_{3}^{(7)}$} & \tabnode{$3C_{3}^{(8)}$} & \tabnode{$1C_{3}^{(1)}$} & \tabnode{$1C_{3}^{(2)}$} \\ \hline
&$1$&$A^2B^2$& $A^2B$ &$A$&$AB^2$&$AB$&$A^2$&$B^2$&$B$&$BAB^2A^2$&$ABA^2B^2$ \\ \hline
~~~$\bm{1_{0,0}}$~~~&     1 & 1&   1 & 1&  1&  1&  1 & 1&  1 & 1 & 1\\
~~~\tabnode{$\bm{1_{0,1}}$}~~~&      1 &$\omega^2$ & $\omega$ &  1 &$\omega^2$& $\omega$ & 1 &$\omega^2$& $\omega$ & 1 & 1\\
		~~~\tabnode{$\bm{1_{0,2}}$}~~~&     1&  $\omega$ & $\omega^2$ & 1 & $\omega$ & $\omega^2$& 1 & $\omega$ & $\omega^2$ & 1&  1\\
		~~~\tabnode{$\bm{1_{1,0}}$}~~~&      1& $\omega^2$ &$\omega^2$ & $\omega$ & $\omega$ &  $\omega$ &$\omega^2$ & 1&  1 & 1&  1\\
		~~~\tabnode{$\bm{1_{1,1}}$}~~~&      1 & $\omega$ & 1 & $\omega$ & 1 &$\omega^2$ &$\omega^2$ &$\omega^2$& $\omega$ & 1 & 1\\
		~~~\tabnode{$\bm{1_{1,2}}$}~~~&     1 & 1 & $\omega$ & $\omega$ &$\omega^2$& 1 &$\omega^2$& $\omega$ & $\omega^2$ & 1 & 1\\
		~~~\tabnode{$\bm{1_{2,0}}$}~~~&     1  &$\omega$ &  $\omega$ & $\omega^2$ &$\omega^2$ &$\omega^2$ & $\omega$ & 1&  1&  1&  1\\
		~~~\tabnode{$\bm{1_{2,1}}$}~~~&     1  &1 &$\omega^2$ &$\omega^2$ & $\omega$ & 1&  $\omega$ & $\omega^2$ & $\omega$ & 1&  1\\
		~~~\tabnode{$\bm{1_{2,2}}$}~~~&     1 &$\omega^2$ & 1& $\omega^2$ & 1&  $\omega$ & $\omega$ & $\omega$ &$\omega^2$& 1 & 1\\
		~~~\tabnode{$\bm{3}$}~~~&     3&  0&  0& 0&  0& 0&  0&  0& 0&  $3 \omega$ &$3\omega^2$\\
		~~~\tabnode{$\bm{\bar{3}}$}~~~&    3  &0&  0& 0&  0&  0 & 0&  0 &0& $3 \omega^2$&  $3 \omega$\\
\end{tabular}
\caption{The character table of $\Delta(27)$, where $\omega=e^{2\pi i/3}$ is the cube root of unity. The arrowed lines show the transformation of the irreducible representations and conjugacy classes of $\Delta(27)$ under the actions of the outer automorphisms $u_S$(blue), $u_T$(also blue) and $u_{K_*}$(green). Note that the action of $u_S$ is identical with $u_T$. The conjugacy classes and representations are left invariant if no arrowed lines are connected with them.  \label{tab:character_Delta27}}
\begin{tikzpicture}[overlay]
\node [above=.3cm,minimum width=0pt] at (1) (c1){};
\node [above=.3cm,minimum width=0pt] at (2) (c2){};
\node [above=.3cm,minimum width=0pt] at (3) (c3){};
\node [above=.3cm,minimum width=0pt] at (4) (c4){};
\node [above=.3cm,minimum width=0pt] at (5) (c5){};
\node [above=.3cm,minimum width=0pt] at (6) (c6){};
\node [above=.3cm,minimum width=0pt] at (7) (c7){};
\node [above=.3cm,minimum width=0pt] at (8) (c8){};
\node [above=.3cm,minimum width=0pt] at (9) (c9){};
\node [above=.3cm,minimum width=0pt] at (10) (c10){};
\node [above=.3cm,minimum width=0pt] at (11) (c11){};
\node [left=.3cm,minimum width=0pt] at (12) (r12){};
\node [left=.3cm,minimum width=0pt] at (13) (r13){};
\node [left=.3cm,minimum width=0pt] at (14) (r14){};
\node [left=.3cm,minimum width=0pt] at (15) (r15){};
\node [left=.3cm,minimum width=0pt] at (16) (r16){};
\node [left=.3cm,minimum width=0pt] at (17) (r17){};
\node [left=.3cm,minimum width=0pt] at (18) (r18){};
\node [left=.3cm,minimum width=0pt] at (19) (r19){};
\node [left=.1cm,minimum width=0pt] at (20) (r31){};
\node [left=.1cm,minimum width=0pt] at (21) (r32){};
\node [right=.2cm,minimum width=0pt] at (12) (r12ri){};
\node [right=.2cm,minimum width=0pt] at (13) (r13ri){};
\node [right=.2cm,minimum width=0pt] at (14) (r14ri){};
\node [right=.2cm,minimum width=0pt] at (15) (r15ri){};
\node [right=.2cm,minimum width=0pt] at (16) (r16ri){};
\node [right=.2cm,minimum width=0pt] at (17) (r17ri){};
\node [right=.2cm,minimum width=0pt] at (18) (r18ri){};
\node [right=.2cm,minimum width=0pt] at (19) (r19ri){};
\node [right=.1cm,minimum width=0pt] at (20) (r31ri){};
\node [right=.1cm,minimum width=0pt] at (21) (r32ri){};
\draw [<->,out=45,in=135,blue!50, very thick,below=1cm] (c2) to (c6);
\draw [<->,out=35,in=145,blue!50, very thick,below=1cm] (c3) to (c5);
\draw [<->,out=45,in=135,blue!50, very thick,below=1cm] (c4) to (c7);
\draw [<->,out=30,in=150,blue!50, very thick,below=1cm] (c8) to (c9);
\draw [<->,out=-45,in=45,blue!50, very thick,below=1cm] (r12ri) to (r13ri);
\draw [<->,out=-40,in=40,blue!50, very thick,below=1cm] (r14ri) to (r17ri);
\draw [<->,out=-50,in=60,blue!60, very thick,below=1cm] (r15ri) to (r19ri);
\draw [<->,out=-40,in=40,blue!50, very thick,below=1cm] (r16ri) to (r18ri);
\draw [<->,out=180,in=180,green!50, very thick,below=1cm] (r31) to (r32);
		\draw [<->,out=215,in=145,green!50, very thick,below=1cm] (r12) to (r13);
		\draw [<->,out=215,in=145,green!50, very thick,below=1cm] (r14) to (r16);
	\draw [<->,out=215,in=145,green!50, very thick,below=1cm] (r17) to (r18);
		\draw [<->,out=45,in=135,green!50, very thick,below=1cm] (c2) to (c9);
		\draw [<->,out=55,in=125,green!50, very thick,below=1cm] (c3) to (c5);
		\draw [<->,out=55,in=125,green!50, very thick,below=1cm] (c6) to (c8);
\draw [<->,out=45,in=135,green!50, very thick,below=1cm] (c10) to (c11);
	\end{tikzpicture}
\end{table}
In the following, we present CG coefficients in the chosen basis. All CG coefficients can be reported in the form $\alpha\otimes\beta$, where $\alpha_{i}$ denotes the elements of the left base vector $\alpha$, and $\beta_{j}$ stands for the elements of the right base vector $\beta$. In the following, we shall adopt the convention $\beta_{[3]}=\beta_{0}\equiv\beta_{3}$. We first report the CG coefficients associated with the singlet representation $\bm{1_{r,s}}$
\begin{equation*}
\bm{1_{r,s}} \otimes \bm{3}  = \bm{3} \,\,\sim\,\, \alpha _1\left(
\begin{array}{c}
\beta_{[1-s]}  \\
\omega^r \beta_{[2-s]}  \\
\omega^{2r}  \beta_{[3-s]}
\end{array}
\right), \qquad
\bm{1_{r,s} }\otimes\bm{\overline{3}}   = \bm{\overline{3}} \,\,\sim\,\, \alpha _1\left(
\begin{array}{c}
\beta_{[1+s]}\\
\omega^r \beta_{[2+s]}\\
\omega^{2r} \beta_{[3+s]}
\end{array}
\right) \,.
\end{equation*}
Finally, for the products of the triplet representations $\bm{3}$ and $\bm{\bar{3}}$, we find
\begin{eqnarray*}
\nonumber &&\hskip-0.5cm \bm{3}\otimes\bm{\overline{3}}= \sum_{r,s=0}^2 \bm{1_{r,s}} \quad \text{with} \quad
\bm{1_{r,s}}:\alpha_1 \beta_{[1-s]} + \omega^{-r}\alpha_2 \beta_{[2-s]} + \omega^{r}\alpha_3 \beta_{[3-s]}\,, \\
&&\hskip-0.5cm \bm{3}\otimes\bm{3}= \bm{\overline{3}_{S_1}} \oplus   \bm{\overline{3}_{S_2} }\oplus \bm{\overline{3}_A } (\bm{\overline{3}}\otimes\bm{\overline{3}}=\bm{3_{S_1}}      \oplus\bm{3_{S_2}}\oplus\bm{3_A }) \quad \text{with}\quad
\left\{\begin{array}{l}
\bm{\overline{3}_{S_1}} (\bm{3_{S_1}}) : \left(
\begin{array}{c}
\alpha_1 \beta_1\\
\alpha_2 \beta_2\\
\alpha_3 \beta_3
\end{array}
\right)\,, \\
\bm{\overline{3}_{S_2}} (\bm{3_{S_2}}) : \left(
\begin{array}{c}
	\alpha_2 \beta_3 + \alpha_3 \beta_2
	\\
	\alpha_3 \beta_1 + \alpha_1 \beta_3
	\\
	\alpha_1 \beta_2 + \alpha_2 \beta_1
\end{array}
\right)\,, \\
\bm{\overline{3}_{A}} (\bm{3_{A}}) : \left(
\begin{array}{c}
	\alpha_2 \beta_3 - \alpha_3 \beta_2
	\\
	\alpha_3 \beta_1 - \alpha_1 \beta_3
	\\
	\alpha_1 \beta_2 - \alpha_2 \beta_1
\end{array}
\right)\,.
\end{array}\right.
\end{eqnarray*}

\section{\label{sec:S3_group}The finite modular group $\Gamma_{2}\cong S_{3}$ and modular forms of level 2 }

The group $\Gamma_{2}\cong S_{3}$ is the permutation group of order $3$ with 6 elements. It can be expressed in terms of two generators $S$ and $T$ which satisfy the multiplication rules~\cite{Ishimori:2010au}
\begin{equation}
S^2=T^2=(ST)^3=1\,.
\end{equation}
The six elements of $\Gamma_{2}\cong S_{3}$  can be divided into three conjugacy classes
\begin{equation}
1C_1=\left\{1 \right\}, \quad 3C_2=\{S,T,TST\},\quad  2C_3=\{ST,TS\}\,,
\end{equation}
where the conjugacy class is denoted by $nC_k$. $k$ is the number of elements belonging to it, and the subscript $n$ is the order of the elements contained in it. The finite modular group $S_{3}$ has two singlet representations $\bm{1}$ and $\bm{1^\prime}$, and one double representation $\bm{2}$. In the present work, we shall work in the basis where the representation matrix of the generator $T$ is diagonal. The representation matrices of generators $S$ and $T$ in three irreducible representations are taken to be
\begin{eqnarray}
\nonumber  \bm{1}&:&~~ \rho_{{\bm 1}}(S)=1, \qquad \rho_{{\bm 1}}(T)=1 \,,  \\
\nonumber  \bm{1^\prime}&:&~~ \rho_{{\bm 1^\prime}}(S)=-1, \qquad \rho_{{\bm 1^\prime}}(T)=-1 \,,  \\
\label{eq:Tp_irre}  \bm{2}&:&~~ \rho_{{\bm 2}}(S)=-\frac{1}{2}
\left(\begin{array}{cc}	1 &~  \sqrt{3} \\
\sqrt{3} &~ -1 \\
\end{array}\right), \qquad \rho_{{\bm 2}}(T)=\left(\begin{array}{cc}
1 &~ 0 \\
0 &~ -1 \\
\end{array}\right) \,.
\end{eqnarray}
The Kronecker products between different irreducible representations can be obtained from the character table
\begin{equation}
\label{eq:S3_KP}
\bm{1}\otimes \bm{1^\prime}=\bm{1^{\prime}} , \qquad  \bm{1^a}\otimes \bm{2}=\bm{2}, \qquad \bm{2}\otimes \bm{2}= \bm{1}\oplus \bm{1^{\prime}}\oplus \bm{2}\,,
\end{equation}
where $a,b=0,1$ and we denote $\bm{1^0}\equiv\bm{1}$ and $\bm{1^1}\equiv\bm{1^\prime}$. We now list the CG coefficients in our working basis. For the product of the singlet $\bm{1^{\prime}}$ with a doublet, we have
\begin{equation}
\bm{1^\prime}\otimes\bm{2}=\bm{2}\,\,\sim\,\,\alpha\left(\begin{array}{c}
 \beta_2 \\
  -\beta_1
\end{array}\right)\,.
\end{equation}
The CG coefficients for the products involving the doublet representation $\bm{2}$ are found to be
\begin{eqnarray*}
\begin{array}{lll}
\bm{2}\otimes\bm{2}=\bm{1}\oplus\bm{1^\prime}\oplus\bm{2},& \qquad\qquad &
\text{with}\qquad \left\{\begin{array}{l}
\bm{1}\;=\alpha_1\beta_1+\alpha_2\beta_2\,,  \\
\bm{1^\prime}=\alpha_1\beta_2-\alpha_2\beta_1\,,  \\
\bm{2}\;=\left(\begin{array}{c}
\alpha_2\beta_2- \alpha_1\beta_1  \\
\alpha_1\beta_2+\alpha_2\beta_1
\end{array}\right)\,.
\end{array}\right. \\
\\[-8pt]
\end{array}
\end{eqnarray*}

In this eclectic approach, the Yukawa couplings are modular forms which are holomorphic functions of the complex modulus $\tau$. In our model, all of the couplings as well as lepton masses must be modular forms of even weights of level 2.  There are two linearly independent modular forms of the lowest non-trivial weight 2. They have been derived in Ref.~\cite{Kobayashi:2018vbk} and  they are explicitly written by use of eta-function as
\begin{eqnarray}
	Y_1(\tau) &=&\frac{i}{4\pi}\left[ \frac{\eta'(\tau/2)}{\eta(\tau/2)}  +\frac{\eta'((\tau +1)/2)}{\eta((\tau+1)/2)}
	- 8\frac{\eta'(2\tau)}{\eta(2\tau)}  \right] ,\nonumber \\
\label{eq:MF_level2_weight2}	Y_2(\tau) &=& \frac{\sqrt{3}i}{4\pi}\left[ \frac{\eta'(\tau/2)}{\eta(\tau/2)}  -\frac{\eta'((\tau +1)/2)}{\eta((\tau+1)/2)}   \right]  ,
\end{eqnarray}
which can be arranged into a doublet $\bm{2}$ of $S_{3}$ and the doublet is taken to be
\begin{equation}
  Y^{(2)}_{\bm{2}}=\left(\begin{array}{c}
Y_1(\tau) \\ Y_2(\tau)
  \end{array}\right)\,.
\end{equation}
The Dedekind function $\eta(\tau)$ in Eq.~\eqref{eq:MF_level2_weight2} is defined as
\begin{equation}
	\eta(\tau)=q^{1/24}\prod_{n=1}^\infty \left(1-q^n \right),\qquad
	q\equiv e^{2 \pi i\tau}\,.
\end{equation}
Then we can obtain that the $q$-expansion of the modular forms $Y_{1,2}(\tau)$ is given by
\begin{eqnarray}
	Y_1(\tau) &=& 1/8+3 q+3 q^2+12 q^3+3 q^4+18 q^5+12 q^6+24 q^7+3 q^8+39 q^9+18 q^{10} \cdots ,\nonumber \\
	Y_2(\tau) &=& \sqrt 3 q^{1/2} (1+4 q+6 q^2+8 q^3+13 q^4+12 q^5+14 q^6+24 q^7+18 q^8+20 q^9 \cdots  ).
\end{eqnarray}
As the expressions of the linearly independent higher weight modular multiplets can be obtained from the tensor products of lower weight modular multiplets. The explicit expressions of the modular multiplets of level $N=2$ up to weight 8 are
\begin{equation}\label{eq:Yw2to8}
	\begin{array}{lll}
		Y^{(4)}_{\bm{1}}
		=\left(Y^{(2)}_{\bm{2}}Y^{(2)}_{\bm{2}}\right)_{\bm{1}}=(Y^{(2)}_{\bm{2},1})^2+(Y^{(2)}_{\bm{2},2})^2\,, & ~
		Y^{(4)}_{\bm{2}}
		=\left(Y^{(2)}_{\bm{2}}Y^{(2)}_{\bm{2}}\right)_{\bm{2}}
=\left(\begin{array}{c}(Y^{(2)}_{\bm{2},2})^2-(Y^{(2)}_{\bm{2},1})^2\\
2 Y^{(2)}_{\bm{2},1} Y^{(2)}_{\bm{2},2}\end{array}\right)\,, \\
		Y^{(6)}_{\bm{1}}
		=\left(Y^{(2)}_{\bm{2}}Y^{(4)}_{\bm{2}}\right)_{\bm{1}} =Y^{(2)}_{\bm{2},1} Y^{(4)}_{\bm{2},1}+Y^{(2)}_{\bm{2},2} Y^{(4)}_{\bm{2},2}, &		
Y^{(6)}_{\bm{1^{\prime}}}
		=\left(Y^{(2)}_{\bm{2}}Y^{(4)}_{\bm{2}}\right)_{\bm{1^{\prime}}}=Y^{(2)}_{\bm{2},1} Y^{(4)}_{\bm{2},2}-Y^{(2)}_{\bm{2},2} Y^{(4)}_{\bm{2},1} \,,\\
		Y^{(6)}_{\bm{2}}
		=\left(Y^{(2)}_{\bm{2}}Y^{(4)}_{\bm{1}}\right)_{\bm{2}}=\left(\begin{array}{c}Y^{(2)}_{\bm{2},1} Y^{(4)}_{\bm{1}}\\
Y^{(2)}_{\bm{2},2} Y^{(4)}_{\bm{1}}\end{array}\right) \,, & ~
		Y^{(8)}_{\bm{1}}
		=\left(Y^{(4)}_{\bm{1}}Y^{(4)}_{\bm{1}}\right)_{\bm{1}}=(Y^{(4)}_{\bm{1}})^2\,, \\
		Y^{(8)}_{\bm{2a}}
		=\left(Y^{(4)}_{\bm{1}}Y^{(4)}_{\bm{2}}\right)_{\bm{2}}=\left(\begin{array}{c}Y^{(4)}_{\bm{1}} Y^{(4)}_{\bm{2},1}\\
Y^{(4)}_{\bm{1}} Y^{(4)}_{\bm{2},2}\end{array}\right)\,, & ~
		Y^{(8)}_{\bm{2b}}
		=\left(Y^{(4)}_{\bm{2}}Y^{(4)}_{\bm{2}}\right)_{\bm{2}}=\left(\begin{array}{c}(Y^{(4)}_{\bm{2},2})^2-(Y^{(4)}_{\bm{2},1})^2\\
2 Y^{(4)}_{\bm{2},1} Y^{(4)}_{\bm{2},2}\end{array}\right) \,.&
	\end{array}
\end{equation}

\section{\label{sec:Delta27-invariant-contractions}The $\Delta(27)$ invariant contractions of two doublets}

In this section, we shall show the contraction results of two $\Delta(27)$ doublet fields $\Phi=(\phi_{1},\phi_{2})^{T}\sim\bm{2_{i,m}}$ and $\Theta=(\theta_{1},\theta_{2})^T\sim\bm{2_{j,n}}$, where  $i,j=1,\cdots 4$ and $m,n=0,1,2$. The $\Delta(27)$ invariant contraction $(\Phi\Theta)_{\bm{\mathcal{R}}}$ requires $i=j$. Without loss of generality, we only consider the assignments of $\Phi$ and $\Theta$ with $m\leq n$. Then there are total 24 possible different assignments for fields $\Phi$ and $\Theta$. The contraction results of the 24 assignments are summarized as follow
\begin{eqnarray}
\nonumber &&n=m \quad : \quad \left\{\begin{array}{l}(\Phi\Theta)_{\bm{1^{0}_{0,0}}}=\phi_{1}\theta_{2} +\phi_{2}\theta_{1} \,, \\
		(\Phi\Theta)_{\bm{1^{1}_{0,0}}}=\phi_{1}\theta_{2} -\phi_{2}\theta_{1} \,,
\end{array}\right. \\
\nonumber&&n=m+1\quad : \quad (\Phi\Theta)_{\bm{2_{0}}}=\left(\begin{array}{c}
\phi_{1}\theta_{2} +\omega^2\phi_{2} \theta_{1} \\
	i \phi_{1}\theta_{2} -i\omega^2\phi_{2} \theta_{1}
\end{array}\right)\,,\\
\label{eq:two_doublet_cont}&&n=m+2 \quad : \quad (\Phi\Theta)_{\bm{2_{0}}}=\left(\begin{array}{c}
	\phi_{1}\theta_{2} +\omega \phi_{2}\theta_{1}  \\
	-i \phi_{1}\theta_{2} +i\omega \phi_{2}\theta_{1}
	\end{array}\right)\,.
\end{eqnarray}
When the two $\Delta(27)$ doublet fields $\Phi=(\phi_{1},\phi_{2})^{T}$ and $\Theta=(\theta_{1},\theta_{2})^T$ transform as $\bm{2_{i,m}}$ and $\bm{2_{i,n}}$ under the EFG $\Delta(27)\rtimes S_{3}$, respectively. The $\Delta(27)$ invariant contraction results of $\Theta^\dagger\Phi$ are given by
\begin{eqnarray}
\nonumber &&n=m \quad : \quad \left\{\begin{array}{l}(\Theta^\dagger\Phi)_{\bm{1^{0}_{0,0}}}=\theta^\dagger_{1}\phi_{1} +\theta^\dagger_{2}\phi_{2} \,, \\
		(\Theta^\dagger\Phi)_{\bm{1^{1}_{0,0}}}=\theta^\dagger_{1}\phi_{1} -\theta^\dagger_{2}\phi_{2} \,,
	\end{array}\right. \\
\nonumber&&n=m+1\quad : \quad (\Theta^\dagger\Phi)_{\bm{2_{0}}}=\left(\begin{array}{c}
	\theta^\dagger_{1}\phi_{1} +\omega^2\theta^\dagger_{2}\phi_{2}  \\
	i \theta^\dagger_{1}\phi_{1} -i\omega^2\theta^\dagger_{2}\phi_{2}
	\end{array}\right)\,,\\
&&n=m+2 \quad : \quad (\Theta^\dagger\Phi)_{\bm{2_{0}}}=\left(\begin{array}{c}
	\theta^\dagger_{1}\phi_{1} +\omega \theta^\dagger_{2}\phi_{2}  \\
	-i \theta^\dagger_{1}\phi_{1} +i\omega \theta^\dagger_{2}\phi_{2}
	\end{array}\right)\,,
\end{eqnarray}
which can be obtained from Eq.~\eqref{eq:two_doublet_cont} through the replacements $\theta_{1}\rightarrow\theta^\dagger_{2}$ and $\theta_{2}\rightarrow\theta^\dagger_{1}$.

\section{\label{sec:EFG_CP_like}The CP-like transformations of EFG $\Delta(27)\rtimes S_{3}$}

In this Appendix, we will follow Refs.~\cite{Chen:2014tpa,Ratz:2019zak} to study the CP-like transformations of the EFG $\Delta(27)\rtimes S_{3}$. We find that the EFG $\Delta(27)\rtimes S_{3}$ has 568 involutory automorphisms whose squared are the trivial identity automorphism. For each involutory automorphism, there exists at least one irreducible representation whose twisted Frobenius-Schur indicator equals to zero. Hence $\Delta(27)\rtimes S_{3}$ is a group of \textbf{Type I} in the terminology of~\cite{Chen:2014tpa,Ratz:2019zak}. One could impose a consistent CP transformation in non-generic setting which contains only a subset of irreducible representations~\cite{Chen:2014tpa,Ratz:2019zak}, although it is impossible to define a proper CP transformation in the context of EFG $\Delta(27)\rtimes S_{3}$ for a generic field content. There are usually more than one possible choices of the CP-like symmetry in non-generic setup for a \textbf{Type I}  group. Given the consistency condition between CP symmetry and modular symmetry in Eq.~\eqref{eq:auto_Kstar_rules}, we find there are only three nontrivial involutory automorphisms for the CP-like transformations of the EFG $\Delta(27)\rtimes S_{3}$:
\begin{eqnarray}
\nonumber u_{K^{*}}&:& A\to A, \quad B\to AB^2A, \quad S\to S^{-1}, \quad T\to T^{-1}\,, \\
\nonumber u^{\prime}_{K^{*}}&:& A\to A, \quad B\to B^2, \quad S\to S^{-1}, \quad T\to T^{-1}\,, \\
u^{\prime\prime}_{K^{*}}&:& A\to A, \quad B\to BAB, \quad S\to S^{-1}, \quad T\to T^{-1}\,.
\end{eqnarray}
The generalized CP transformation corresponding to $u_{K^{*}}$ has been analyzed in section~\ref{subsec:gCP}. Now we proceed to discuss the remaining $u^{\prime}_{K^{*}}$ and $u^{\prime\prime}_{K^{*}}$. The automorphism  $u^{\prime}_{K^{*}}$ acts on the irreducible representations of the EFG $\Delta(27)\rtimes S_{3}$ as follow
\begin{eqnarray}
\nonumber u^{\prime}_{K^{*}}&:& \bm{1^{a}_{0,0}}\leftrightarrow \bm{1^{a}_{0,0}}, \quad \bm{2_{0}}\leftrightarrow \bm{2_{0}}, \quad \bm{2_{1,0}}\leftrightarrow \bm{2_{1,0}}, \quad \bm{2_{1,1}}\leftrightarrow \bm{2_{1,2}},   \\
\label{eq:uKpstar-rep} && \bm{2_{2,m}}\leftrightarrow \bm{2_{4,m}}, \quad \bm{2_{3,m}}\leftrightarrow \bm{2_{3,m}}, \quad \bm{3^{a}}\leftrightarrow \bm{\bar{3}^{a}}, \quad \bm{6}\leftrightarrow\bm{\bar{6}}\,.
\end{eqnarray}
Recalling that the representations $(\bm{3^{0}},\bm{\bar{3}^{0}})$, $(\bm{3^{1}},\bm{\bar{3}^{1}})$ and $(\bm{6},\bm{\bar{6}})$ are the complex conjugate of each other, and the other representations of $\Delta(27)\rtimes S_{3}$ are self-conjugate. Therefore we can see that $u^{\prime}_{K^{*}}$ maps the irreducible representations $\bm{1^{a}_{0,0}}$, $\bm{2_{0}}$,  $\bm{2_{1,0}}$, $\bm{2_{3,m}}$, $\bm{3^{a}}$, $\bm{\bar{3}^{a}}$, $\bm{6}$ and $\bm{\bar{6}}$ to their complex conjugates. If a model only contains a subset of these irreducible representations, one can consistently impose the CP-like transformation corresponding to $u^{\prime}_{K^{*}}$ to be a symmetry, and thus guarantee CP conservation. In our working basis, the CP transformation matrices $\rho_{\bm{r}}( K^{*})$ for the automorphism $u^{\prime}_{K^{*}}$ are determined to be
\begin{eqnarray}
\nonumber  &&\rho_{\bm{1^{a}_{0, 0}}}(K^{*})=1,\quad \rho_{\bm{2_{0}}}(K^{*})=\rho_{\bm{2_{1,0}}}(K^{*})=\mathbb{1}_{2}\,, \quad
\rho_{\bm{2_{3,m}}}(K^{*})=\left(
\begin{array}{cc}
0 & 1 \\
1 & 0 \\
\end{array}
\right)\,, \\
\label{eq:gCP_transf2}  && \rho_{\bm{3^{a}}}(K^{*})=
\rho_{\bm{\bar{3}^{a}}}(K^{*})=\mathbb{1}_{3}\,,  \quad
\rho_{\bm{6}}(K^{*})=	\rho_{\bm{\bar{6}}}(K^{*})=\mathbb{1}_{6}\,.
\end{eqnarray}
Analogously the action of $u^{\prime\prime}_{K^{*}}$ on the irreducible representations of $\Delta(27)\rtimes S_{3}$ is given by
\begin{eqnarray}
\nonumber u^{\prime\prime}_{K^{*}}&:& \bm{1^{a}_{0,0}}\leftrightarrow \bm{1^{a}_{0,0}}, \quad \bm{2_{0}}\leftrightarrow \bm{2_{0}}, \quad \bm{2_{1,0}}\leftrightarrow \bm{2_{1,0}}, \quad \bm{2_{1,1}}\leftrightarrow \bm{2_{1,2}},  \\
\label{eq:uKppstar-rep}&& \bm{2_{2,m}}\leftrightarrow \bm{2_{3,m}}, \quad \bm{2_{4,m}}\leftrightarrow \bm{2_{4,m}}, \quad \bm{3^{a}}\leftrightarrow \bm{\bar{3}^{a}}, \quad \bm{6}\leftrightarrow\bm{\bar{6}}\,.
\end{eqnarray}
Consequently the CP symmetry corresponding to $u^{\prime\prime}_{K^{*}}$ can be properly implemented in a model containing only a subset of the irreducible representations $\bm{1^{a}_{0,0}}$, $\bm{2_{0}}$,  $\bm{2_{1,0}}$, $\bm{2_{4,m}}$, $\bm{3^{a}}$, $\bm{\bar{3}^{a}}$, $\bm{6}$ and $\bm{\bar{6}}$. Solving the consistency condition, the CP transformation matrices are fixed as
\begin{eqnarray}
\nonumber  &&\rho_{\bm{1^{a}_{0, 0}}}(K^{*})=1,\quad \rho_{\bm{2_{0}}}(K^{*})=\rho_{\bm{2_{1,0}}}(K^{*})=\mathbb{1}_{2}\,, \quad
\rho_{\bm{2_{4,m}}}(K^{*})=\left(\begin{array}{cc}
0 & 1 \\
1 & 0 \\
\end{array}
\right)\,, \\
\nonumber  && \rho_{\bm{3^{a}}}(K^{*})=\rho^{*}_{\bm{\bar{3}^{a}}}(K^{*})=\frac{1}{\sqrt{3}}\left(
\begin{array}{ccc}
\omega & 1 & 1 \\
1 & \omega & 1 \\
1 & 1 & \omega \\
\end{array}
\right)\,, \\
\label{eq:gCP_transf3} && \rho_{\bm{6}}(K^{*})=\rho^{*}_{\bm{\bar{6}}}(K^{*})=\frac{1}{\sqrt{3}}\left(
\begin{array}{ccc}
	\omega\mathbb{1}_{2} & \mathbb{1}_{2} & \mathbb{1}_{2} \\
	\mathbb{1}_{2} & \omega \mathbb{1}_{2} & \mathbb{1}_{2} \\
	\mathbb{1}_{2} & \mathbb{1}_{2} & \omega \mathbb{1}_{2} \\
\end{array}
\right)\,.
\end{eqnarray}
Since a flavon $\varphi$ in the representation $\bm{2_{2,0}}$ is introduced in our model, one cannot impose the CP symmetry corresponding to the automorphisms $u^{\prime}_{K^{*}}$ and $u^{\prime\prime}_{K^{*}}$ in the model. The EFG models with the CP-like symmetry $u^{\prime}_{K^{*}}$ or $u^{\prime\prime}_{K^{*}}$ could be constructed as well, we expect that the constraints on the couplings would be different from Eq.~\eqref{eq:CP_cons_Mpsi}, Eq.~\eqref{eq:CP_cons_Mpsi2} and Eq.~\eqref{eq:CP_cons_Mpsip}.

\end{appendix}

\providecommand{\href}[2]{#2}\begingroup\raggedright\endgroup

\end{document}